\documentclass[a4paper,fleqn,usenatbib]{mnras}
\usepackage{newtxtext,newtxmath}
\usepackage[T1]{fontenc}
\usepackage{ae,aecompl}
\usepackage{float}
\usepackage{color}

\usepackage{graphicx}	% Including figure files
\usepackage{amsmath}	% Advanced maths commands
\usepackage{amssymb}	% Extra maths symbols
\usepackage{xcolor}
\usepackage{etoolbox}
\makeatletter
\patchcmd\@combinedblfloats{\box\@outputbox}{\unvbox\@outputbox}{}{\errmessage{\noexpand patch failed}}
\makeatother

\title[Strongly magnetized disks]{Strongly magnetized accretion disks: structure and accretion from global magnetohydrodynamic simulations}

\author[Mishra et al.]{
Bhupendra Mishra $^{1}$ \thanks{E-mail: bhupendra.mishra@jila.colorado.edu}, Mitchell C. Begelman $^{1,2}$, Philip J. Armitage $^{3,4}$, Jacob B. Simon $^{1,5,6}$
\\
$^{1}$ JILA, University of Colorado and National Institute of Standards and Technology, 440 UCB, Boulder, CO 80309-0440, USA \\
$^{2}$ Department of Astrophysical and Planetary Sciences, University of Colorado, 391 UCB, boulder, CO 80309-0391, USA \\
$^{3}$ Center for Computational Astrophysics, Flatiron Institute, 162 Fifth Avenue, New York, NY 10010, USA \\
$^{4}$ Department of Physics and Astronomy, Stony Brook University, Stony Brook, NY 11790, USA \\
$^{5}$ Department of Space Studies, Southwest Research Institute, Boulder, CO 80302, USA \\
$^{6}$ Department of Physics and Astronomy, Iowa State University, Ames, IA, 50010, USA\\
}

\begin{document}
\label{firstpage}
\pagerange{\pageref{firstpage}--\pageref{lastpage}}
\maketitle

% Abstract of the paper
\begin{abstract}
We use global magnetohydrodynamic simulations to study the influence of net vertical magnetic fields on the structure of geometrically thin ($H/r \approx 0.05$) accretion disks in the Newtonian limit. We consider initial mid-plane gas to magnetic pressure ratios $\beta_0 = 1000,\, 300$ and $100$, spanning the transition between weakly and strongly magnetized accretion regimes. We find that magnetic pressure is important for the disks' vertical structure in all three cases, with accretion occurring at $z/R\approx 0.2$ in the two most strongly magnetized models. The disk midplane shows outflow rather than accretion. Accretion through the surface layers is driven mainly by stress due to coherent large scale magnetic field rather than by turbulent stress. Equivalent viscosity parameters measured from our simulations show similar dependencies on initial $\beta_0$ to those seen in shearing box simulations, though the disk midplane is not magnetic pressure dominated even for the strongest magnetic field case. Winds are present but are not the dominant driver of disk evolution. Over the (limited) duration of our simulations, we find evidence that the net flux attains a quasi-steady state at levels that can stably maintain a strongly magnetized disk. We suggest that geometrically thin accretion disks in observed systems may commonly exist in a  
magnetically ``elevated'' state, characterized by non-zero but modest vertical magnetic fluxes, with potentially important implications for disk phenomenology in X-ray binaries (XRBs) and active galactic nuclei (AGN).
\end{abstract}

\section{Introduction}
\label{sec:intro}
 The $\alpha$-viscosity prescription for accretion disks \citep{Shakura73} --- which sets the vertically integrated radial stress proportional to pressure --- assumes that angular momentum transport is driven by turbulence and related to the magnetic field. The physical basis for this prescription in the magnetorotational instability \citep[MRI:][]{Balbus91,Balbus98} seems well-supported by simulations. However, geometrically thin disk models based on the $\alpha$-prescription suffer numerous failures in matching the phenomenology of disks in accreting systems such as cataclysmic variables (CVs), XRBs and AGN. For example, $\alpha$-models were found to be thermally \citep{Shakura76, Piran78} and viscously \citep{Lightman74} unstable  in the inner, radiation pressure-dominated regions of luminous disks, yet observations of black hole XRBs indicate stable accretion all the way to the innermost radius.  In CVs and XRBs, spectral signatures indicate disks that are hotter and geometrically thicker than predicted by the standard theory \citep[and references therein]{Begelman07}. 
 
 The standard thin disk model also fails to support the expected accretion flows around AGN. The outer regions of thin AGN disks are expected to be gravitationally unstable, fragmenting into clumps that trigger star formation. This curtails accretion and hence fails to explain not only the fueling of AGN but also the growth of supermassive black holes \citep{Koly80,Shosman87,Goodman03}. More directly, observations of AGN disk sizes, spectra and variability timescales also point to hotter and geometrically thicker disks in the inner regions \citep{Dexter19}.  
 
 These difficulties can be ameliorated and perhaps resolved if accretion disks are geometrically thicker and faster-accreting than predicted by the standard \cite{Shakura73} theory.  The development of strong toroidal ($B_\phi$) magnetic fields, amplified by the dynamo process associated with MRI, provides a plausible way to obtain such thickening by exerting magnetic pressure in the vertical direction.  A strongly magnetized disk, with a magnetic pressure larger than the gas or radiation pressure, would have a lower density than a standard thin disk with the same accretion rate, thus avoiding gravitational instability \citep{Pariev03,Begelman07,Gaburov12}.  Lower densities could increase the color correction \citep{Blaes06}, leading to higher temperatures, while geometric thickening could increase the degree of radiative reprocessing (also contributing to higher temperatures) and accretion inflow speeds. Finally, the decoupling of radiative transport, dissipation and vertical pressure support would eliminate the thermal-viscous instability \citep{Begelman07, Sadowski16}.  

%===========================================================================================
MRI-driven turbulence tends to amplify the $\phi$-component of the magnetic field more than the other components, but when the net flux of the disk is negligible, the saturated pressure of $B_\phi$ remains small compared to the background pressure. However, this changes if the MRI is seeded by a net poloidal (vertical) magnetic flux of sufficient magnitude. \citet{Hawley95} performed three-dimensional local (shearing box) MHD simulations of the growth and saturation of MRI seeded by   poloidal magnetic field of different strengths. They found that angular momentum transport was dominated by magnetic (Maxwell) stresses rather than Reynolds stress, and characterized by an effective $\alpha$ that increases with increasing initial magnetic field strength. Recent stratified shearing box simulations with higher poloidal fluxes extend the trend measured by \citet{Hawley95} to ratios of gas to magnetic pressure as low as $\beta_0 = 100$  \citep{Bai13} and $\beta_0 = 10$ \citep{Greg16}.  These studies find that values of $\beta_0 \lesssim 1000$ lead to a strongly magnetized disk midplane and regions of large Maxwell stress extending to several scale heights on either side.

Quasi-periodic alternations in the disk's large scale magnetic field (often referred to as a ``dynamo") were observed in vertically stratified local shearing box simulations \citep{Axel95}, and are characteristic of the MRI.  In such a dynamo process, large scale toroidal magnetic fields are generated which show a ``butterfly diagram'' with even parity across the disk midplane and flips in orientation every $\sim 30$ local orbital periods. Although the dynamo develops even with negligible poloidal flux, the characteristics of the field reversals are governed by the initial magnetic field strength. A stronger net poloidal magnetic field delays the reversal of $B_\phi$, leading to field structures that resemble long-lived global patterns rather than turbulence  \citep{Bai13,Greg16}.

Early global simulations of unstratified (cylindrical) disks that included a locally non-zero vertical field yielded moderately large $\alpha$ values $\sim 0.1$ \citep{Armitage98}, qualitatively in accord with local results. Most stratified simulations, 
however, have focused on the zero net flux regime \citep{Mishra16,Sadowski16,Hogg18}. Simulations with a significant net field require a smaller time-step to follow regions with low plasma $\beta$, and also require careful treatment of the inner boundary to avoid numerical artefacts. With a careful choice of the inner boundary condition, \citet{ZhuStone} were able to evolve a fully global numerical MHD model for two different thermal scale heights ($H/r = 0.1$ and $H/r = 0.05$). \citet{ZhuStone} used seed magnetic fields with $\beta_0 =10^4$ and $1000$, with the stronger field case barely (based on local expectations) reaching the boundary of the strongly magnetized regime.

In this paper, we use global simulations to study disks whose net vertical flux is chosen (based on local results) to span the transition from weak to strong magnetic field strengths. Our goal is to test whether the local simulation results carry over to the global regime, and hence whether strongly magnetized disks are potentially relevant to the open problems discussed above.  We ran a weak magnetic field case simulation ($\beta_0 = 10^3$) for 23 orbits (at a fiducial radius $R = 1$). Intermediate ($\beta_0 = 300$) and strong ($\beta_0 = 100$) magnetic field cases are run for 23 and 50 orbits, respectively. We note that \citet{ZhuStone} attempted a model similar to our strong field simulation, but observed extremely rapid mass loss in that case. We find that this problem can be alleviated with a change to the assumed radial temperature structure of the disk, allowing physical results to be obtained in the strong field case.  

The paper is organized as follows. In Section ~\ref{sec:setup}, we describe the numerical setup of the disk which includes the hydrodynamical disk profile and magnetic field configurations. In Section ~\ref{sec:results} we present detailed findings from our numerical simulations. The results are categorized into vertical structure (Section~\ref{sec:vertstructure}), radial structure (Section~\ref{sec:radstructure}), wind properties (Section~\ref{sec:inflowoutflow}), magnetic flux evolution (Section~\ref{sec:fluxBtime}) and comparison with local (Section~\ref{sec:localsim}) shearing box simulations and other global (Section~\ref{sec:globalsim}) simulations. In Section~\ref{sec:discussion}, we conclude and discuss the astrophysical applications and relevance of our numerical model.
%==============================================================================
\begin{figure*}
\centering
\includegraphics[width=1.3\columnwidth]{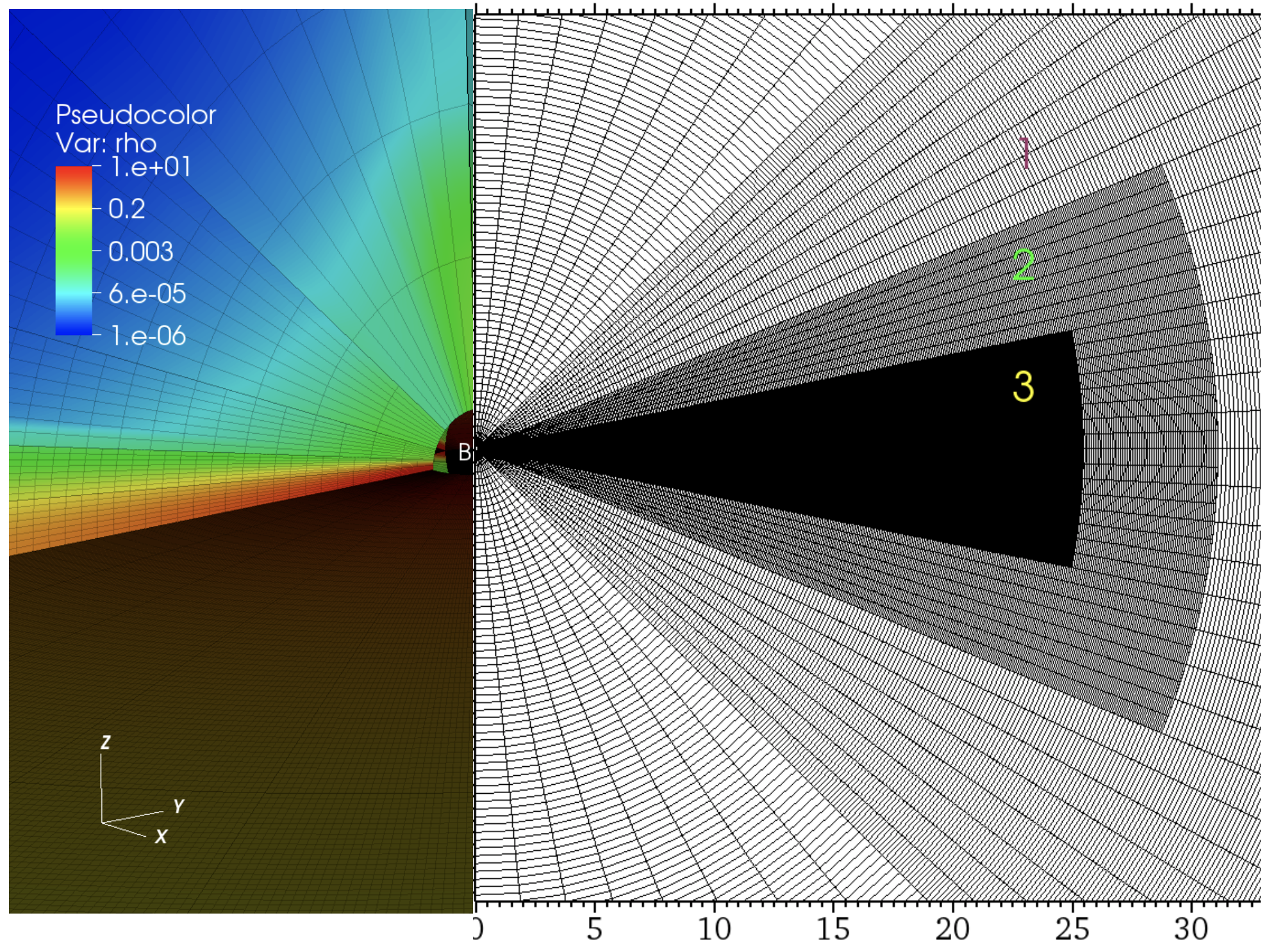}
\caption{Initial configuration showing the three levels of static mesh refinement and density profile. The left panel (with colorbar) shows a 3D view with the density profile following Eq.~\ref{eq:profile_den} and the right panel shows the grid refinement levels numbered 1,2,3 respectively. The simulation domain extends to $R_\mathrm{max} = 100$ but the grid in this plot is shown only out to $R=32$ to highlight the number of refinement levels. The sub-domains symmetrically covered by each refinement level are defined in Section~\ref{sec:setup}.}
\label{fig:initial_grid}
\end{figure*}

%==============================================
%==============================================

\section{Numerical Setup}
\label{sec:setup}
We model the accretion disk using the Athena++ \citep{White16} code in the ideal MHD limit. Athena++ uses a higher-order Godunov scheme and employs constrained transport (CT) to maintain divergence-free evolution of the magnetic field. We use the HLLD solver implemented in Athena++. Our initial conditions and methods are generally similar to those used by \citet{ZhuStone}, with the main difference being that we model a disk with a radially constant disk aspect ratio ($H/r$, where $H$ is the thermal scale height of the disk). The density profile of the disk in hydrostatic equilibrium in cylindrical coordinates is given by 
\begin{equation}
\label{eq:profile_den}
\rho(R,z) = \rho(R,z=0)\exp\left[\frac{GM}{c^2_s}\left(\frac{1}{\sqrt{R^2 + z^2}} - \frac{1}{R}\right)\right],
\end{equation}
where the initial midplane density profile is
\begin{equation}
\rho(R,z=0) = \rho(R_0,z=0)\left(\frac{R}{R_0}\right)^p.
\end{equation}
The initial temperature is taken to be constant on cylinders,
\begin{equation}
T(R,z) = T(R_0)\left(\frac{R}{R_0}\right)^q ,
\end{equation}
and is maintained constant throughout the simulation using the same methods as in \citet{ZhuStone}.
The initial velocity profile in the disk has only an azimuthal component given by
\begin{equation}
v_\phi(R,z) = v_K\left[(p+q)\left(\frac{c_s}{v_{\phi,k}}\right)^2 + 1 + q - \frac{qR}{\sqrt{R^2 + z^2}}\right]^{1/2},
\end{equation}
where $R$ and $z$ are cylindrical radius and height.
%===========================================================================================
We choose $p = -1.5$ and $q = -1$, in contrast to the values $p=-2.25$ and $q=-0.5$ chosen by \citet{ZhuStone}. We are interested in geometrically thin disks and choose an initial disk aspect ratio of $H/r = 0.05$. Note that we wrote the disk profile in cylindrical coordinates for simplicity but the simulations are performed in spherical polar coordinates. 

We maintain effectively isothermal conditions by rapidly relaxing the temperature to a pre-specified value at each cylindrical radius.
The cooling rate is computed by
\begin{equation}
    \frac{dE}{dt} = -\frac{E - c_\nu\rho T_0}{t_\mathrm{cool}},
\end{equation}
where $E$ is internal energy per unit volume and $c_\nu$ is heat capacity per unit mass \citep{Zhu15}.  The rapid cooling time is assumed to be the numerical time step of the simulation. We remove any excess internal energy in the simulation and maintain the same temperature as its initial value at each cylindrical radius. 

Strong magnetic field leads to  small time steps in the inner polar regions of the disk. In order to maintain a reasonable time step we used the radial power-law density profile of floor values prescribed in Eq. 10 of \citet{ZhuStone}:
\begin{equation}
%\[
    \rho_\mathrm{fl}=\left\{
                \begin{array}{ll}
                  \rho_\mathrm{fl,0}\left(\frac{R}{R_0}\right)^p\left(\frac{1}{z^2}\right)\,\, \mathrm{if}\,\,R > r_\mathrm{min}\\
                  \rho_\mathrm{fl,0}\left(\frac{r_\mathrm{min}}{R_0}\right)^p\left(\frac{1}{z^2}\right)\,\, \mathrm{if}\,\,R < r_\mathrm{min}\,\,\mathrm{and}\,\,r>3r_\mathrm{min}\\
                  \rho_\mathrm{fl,0}\left(\frac{R}{R_0}\right)^p\left(\frac{1}{z^2}\right)\left(5 - 2\frac{r-r_\mathrm{min}}{r_\mathrm{min}}\right)\left(4\frac{r_\mathrm{min}-R}{r_\mathrm{min}}+1\right) \\
                  \mathrm{if}\,\, r<r_\mathrm{min}\,\,\mathrm{and}\,\, r<3r_\mathrm{min}
                \end{array}
              \right.
%  \]
\end{equation}
where, $\rho_\mathrm{fl,0}=10^{-9}$ is a constant. The simulation domain extends from $r_\mathrm{min} = 0.1$ to $r_\mathrm{max} = 100$ in code units. The purpose of keeping the outer boundary far away is to avoid any effects of the outer boundary conditions on the simulation region of interest. The polar and azimuthal domains extend from $\theta_\mathrm{min} = 0$ to $\theta_\mathrm{max} = \pi$ and $\phi_\mathrm{min} = 0$ to $\phi_\mathrm{max} = 2\pi$. The radial grid is evenly spaced in logarithmic space. We use the static mesh refinement (SMR) feature of Athena++ to improve the resolution of the dense regions of the disk. We add three levels of static mesh refinement over a base resolution of $256\times 32\times 128$, giving an effective resolution of $2048\times 256\times1024$.

%========================================================================================
\begin{figure*}
\centering
\includegraphics[width=2\columnwidth]{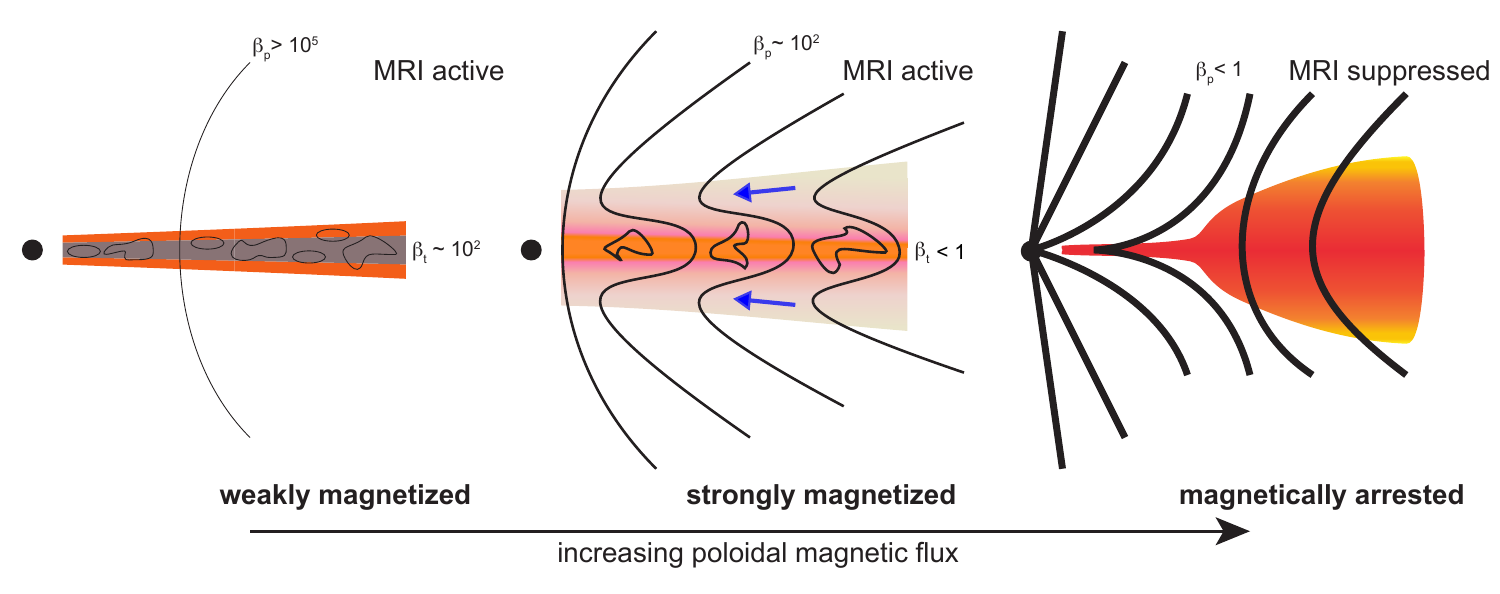}
\caption{Schematic diagram of various accretion regimes. The leftmost  image represents weakly magnetized disks. $\beta_p$ and $\beta_t$ correspond to poloidal and total $\beta$ (which is approximately the same as $\beta$ associated with the toroidal magnetic field). The middle panel is the case we are interested in. The pinching in of the magnetic field on the surface of the disk shows schematically where accretion is occurring. The rightmost panel shows magnetically arrested disks \citep{Narayan03,McKinney12}. Curves illustrate the poloidal magnetic field lines, which can be compared with the actual global field structure in Fig.~\ref{fig:streamline}. The disk thickness variation between the left and center images mimics the effect of strong magnetic pressure support. The left and middle panels are MRI active whereas the rightmost panel is MRI suppressed.}
\label{fig:sketch}
\end{figure*}
%========================================================================================
The static mesh refinement levels (shown in Fig.~\ref{fig:initial_grid}) are applied as follows: all refinement levels extend to the minimum radius $r = 0.1$, which is the inner boundary of the simulation domain. The outer radial limits for the first, second and third refinement level are at $r = 40,\,30,\,20$, respectively. The refinement domains in $\theta$ are applied symmetrically (Fig.~\ref{fig:initial_grid}) about the equator
but Athena++ requires slightly different minimum and maximum values of $\theta$ for best mesh-block configuration of the simulation domain. This leads to the following numbers for $\theta$. The first refinement level extends from $\theta = 0.8$ to $1.8$, the second refinement level extends from $\theta = 1.2$ to $1.77$ and the third refinement level covers the narrow domain of $\theta = 1.4$ to $1.6$ (all in radians). We use an outflowing boundary condition radially for gas and for magnetic field. The azimuthal boundary condition is periodic. In order to avoid significant flux loss from the polar region we use the polar boundary condition described in \citet{ZhuStone}. The polar boundary condition in a full $2\pi$ azimuthal domain acts in a way that assigns cell- and face-centered values into the ghost cell in $\theta$ using the opposite active cell in $\phi$.

The use of an effectively radially isothermal equation of state (and the neglect of physics such as radiative transport) means that the simulations are scale-free. We report all results in code units. We take $G=M=1$ and report densities and velocities in terms of the midplane density and Keplerian velocity at $R=1$.  
%================================================
\subsection{Magnetic Field}
\label{sec:magfield}
As illustrated in Fig.~\ref{fig:sketch}, we can identify three regimes of disk magnetization, all of which are potentially of astrophysical interest. Weakly magnetized disks are those whose poloidal field, even if non-zero, is weak enough that the properties of the MRI are essentially unaltered from the strictly zero net flux limit. Strongly magnetized disks, the subject of this study, are those where the poloidal field, via its ability to source larger toroidal fields, is large enough to affect disk structure qualitatively in the vertical direction, but not so large as to quench MRI and dominate the radial force balance. Finally, magnetically arrested disks
\citep[MAD;][]{Narayan03,McKinney12} occur when the poloidal field on its own is strong enough to dominate the disk dynamics. 

The results of local simulations \citep{Greg16} imply that the transition between weak and strong magnetization occurs for initial values of (poloidal) $\beta = P_{\rm gas} / P_{\rm b}$ in the range between $10^3$ and $10^2$. We therefore seed the initial disk with a purely vertical magnetic field with a uniform ratio of gas pressure to magnetic pressure as a function of radius at the disk midplane. To ensure an initially divergenceless magnetic field, we initialize the magnetic field via the magnetic vector potential defined in \citet{ZhuStone},
\begin{equation}
A_\phi = \frac{B_0}{R_0^m}\frac{R^{m+1}}{m+2} + \frac{B_0 r^{m+2}_\mathrm{min}}{R^m_0}\left(\frac{1}{2} - \frac{1}{m+2}\right),
\end{equation}
where $m = (p+q)/2$ with $p = -1.5$ and $q = -1$. We adopt three values of initial midplane $\beta = P_\mathrm{gas}/P_\mathrm{b}$, corresponding to  weak, intermediate and strong initial magnetic field strength cases with initial $\beta_0 = 1000\, , 300$ and $100$, respectively. We add a sinusoidal component to the initial vertical magnetic field to suppress the growth of channel flows in the strongly magnetized cases. A model similar to our $\beta_0 = 1000$ case (although using a flaring disk profile) has been studied by \citet{ZhuStone} and serves as a point of comparison. 
%=========================================================================================
\section{Results}
\label{sec:results}
% Qualitatively, we find that the simulations support the following conclusions:
% \begin{itemize}
%     \item Elevated accretion or surface accretion occurs at $z/r \approx 0.2$, and it is primarily driven by coherent rather than turbulent Maxwell stresses.
%     \item Several key results of local shearing box simulations are reproduced. In particular the scaling of $\alpha$ with $\beta_0$ reported in local shearing box simulations is recovered.
%     \item Logarithmic spiral structures, with large density contrasts, develop in the strongly magnetized cases.
%     \item Weak winds that carry away only a small fraction of the mass and angular momentum associated with the accretion flow.  
%     \item Evidence that net magnetic flux is neither being accumulated nor lost by the inner regions of the flow, but instead attains an approximate steady-state.
% \end{itemize}
%============================================

We first define the post-processing steps that we employ. An azimuthally averaged quantity $a$ at fixed radius is defined by 
\begin{equation}
    \langle a(R,z)\rangle_\phi = \frac{1}{2\pi} {\int^{2\pi}_0 a(R,z) d\phi}.
\end{equation}
Similarly, a time averaged quantity is defined by 
\begin{equation}
    \langle a \rangle_t = \frac{1}{(t_f - t_i)}\int^{t_f}_{t_i} a dt
 \end{equation}
 where the time averaging is performed between $t_i$ and $t_f$. We ran the $\beta_0 =100$ case for $50$ orbits at $R=1$, while the total runtime of the $\beta_0 =1000$ and $\beta_0 = 300$ simulations is 23 orbits at the same radius. We typically present time averages between $t_i = 19$ and $t_f = 23$. All the vertical profiles reported are azimuthally averaged at $R = 1$ \citep[similar to][]{ZhuStone}. The radial and polar Reynolds and Maxwell stresses are defined by,
 \begin{eqnarray}
     T_{\mathrm{r}\phi\,,\mathrm{Rey}} &=& \rho v_\mathrm{r}\delta v_\phi \\
     T_{\mathrm{r}\phi\,,\mathrm{Max}} &=& -B_\mathrm{r} B_\phi,
 \end{eqnarray}
and,
\begin{eqnarray}
     T_{\theta\phi\,,\mathrm{Rey}} &=& \rho v_\theta\delta v_\phi \\
     T_{\theta\phi\,,\mathrm{Max}} &=& -B_\theta B_\phi,
 \end{eqnarray}
respectively. The Maxwell stress defined above is normalized with a factor of $4\pi$ to be consistent with the Athena++ definition of the magnetic field strength. 
Here $B_r$, $B_\theta$ and $B_\phi$ correspond to components of the magnetic field in spherical polar coordinates, and $\delta v_\phi$ corresponds to velocity fluctuations defined by $\delta v_\phi = (v_\phi - \langle v_\phi \rangle _\phi)$, where $\langle v_\phi \rangle_\phi$ is azimuthally averaged toroidal velocity.

The total radial Maxwell stress $T_{r\phi\,,\mathrm{Max}}$ can be decomposed into coherent and turbulent components. % We compute azimuthally averaged vertical profile these individual components as follows. 
The coherent component is defined by,
\begin{equation}
   \langle T^\mathrm{Coh}_{r\phi\,,\mathrm{Max}}\rangle_\phi = - \langle B_\mathrm{r}\rangle_\phi\langle B_\phi \rangle_\phi
\label{eq:coherentM}
\end{equation}
and the turbulent component is,
\begin{equation}
   \langle T^\mathrm{Turb}_{r\phi\,,\mathrm{Max}}\rangle_\phi = \langle -B_\mathrm{r}B_\phi \rangle_\phi + \langle B_\mathrm{r}\rangle_\phi \langle B_\phi\rangle_\phi.
\label{eq:turbulentM}
\end{equation}
The viscosity parameter $\alpha$ is defined as the ratio of the volume averaged total radial stress (including turbulent and coherent components) to the similarly averaged gas pressure 
%\jsays{I think it may be better to define $\alpha$ as the turbulent stress only.  Not only does this agree with the referee, but it will be much less confusing to the reader.  Is there any problem in redefining things in this way?} \psays{Second Jake here, could do it either way but keeping $\alpha$ as turbulent only is arguably better},
\begin{equation}
    \alpha = \frac{\langle T_{\mathrm{r}\phi} \rangle_v}{\langle P_\mathrm{gas}\rangle_v}.
    \label{eq:alpha}
\end{equation}
We distinguish this from the ``Shakura-Sunyaev" $\alpha_{\rm SS}$, 
\begin{equation}
    \alpha_\mathrm{SS} = \frac{\langle T^\mathrm{Turb}_{r\phi\,,\mathrm{Max}}\rangle_v}{\langle P_\mathrm{gas}\rangle_v}.
    \label{eq:turbalpha}
\end{equation}
which is a similar ratio based only on the turbulent part of the stress, i.e. (Eq.~\ref{eq:turbalpha}). We will see later that the coherent stress dominates the accretion in magnetically dominated disks, leading to $\alpha \gg \alpha_{\rm SS}$ (see Sect 3.1.2).

%===========================================================================================
\begin{figure*}
\centering
\includegraphics[width=0.65\textwidth]{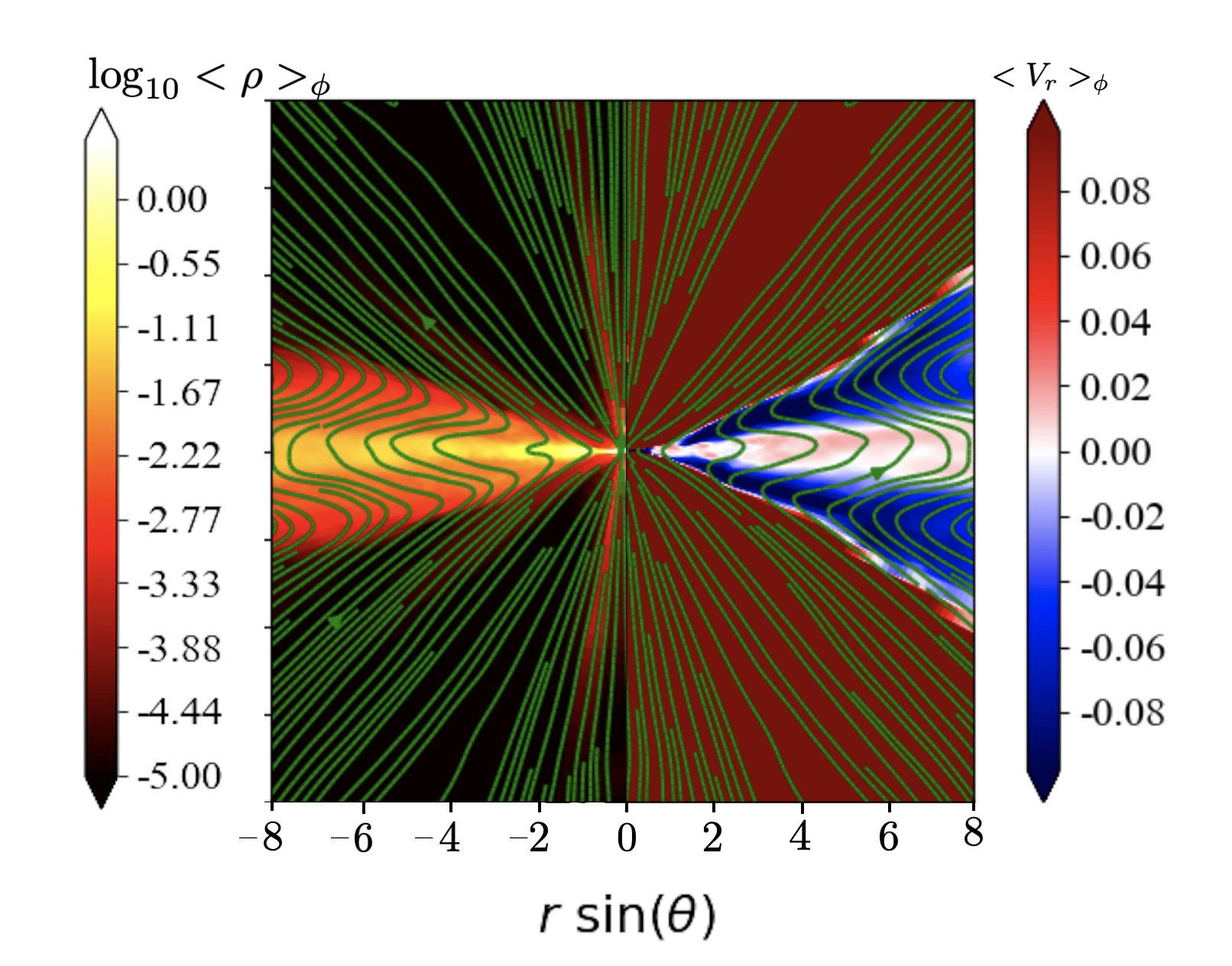}
\caption{Visualization of the strongly magnetized case ($\beta_0 = 100$) at $t = 50$ orbital periods (measured at $R =1$) to show the pinching in of the magnetic field in accreting regions. Green lines show the azimuthalliy averaged poloidal magnetic field. The left and right panels of the image show azimuthally averaged density (in units of density at $R = 1$) and radial velocity (in units of Keplerian velocity at $R =1$), respectively. Blue regions in the right panel correspond to accretion and red regions to outflows.}
\label{fig:streamline}
\end{figure*}
%============================================================
\subsection{Vertical Structure}
\label{sec:vertstructure}
The basic structure of the simulated disks is shown in Fig.~\ref{fig:streamline} for the most strongly magnetized disk with $\beta_0 =100$. This time slice was taken at $t=50$ orbits at $R=1$, at which time the disk and its magnetic field structure had achieved a quasi-steady state (inflow equilibrium) inside $R \approx 3$. Three regions of the flow can be distinguished. Near the disk midplane the magnetic pressure is relatively weak, because of the toroidal field reversal that must occur there when the flow is threaded by net vertical flux. The gas near $z=0$ is outflowing, accompanied by a pushing or pinching of the poloidal magnetic field outwards. Above the midplane, strong accretion occurs in magnetic pressure supported regions, which lie at $z/r \approx 0.2-0.4$ in our models. The poloidal magnetic field is pinched radially inward in this zone of the flow. Finally, the polar or atmosphere regions above the main body of the disk exhibit outflow. Due to the low density in this region (approximately six orders of magnitude lower compared to the density in accreting regions) accretion greatly dominates   outflow, overall.

The primary finding of surface or elevated accretion is consistent with previous studies based on global numerical simulations \citep{Suzuki14,ZhuStone, Jiang19}. However, it is physically different from the structures that had been suggested on the basis of local simulations \citep{Bai13,Greg16} and analytic models \citep{Begelman15}. Although similar values of $\beta_0$ trigger the onset of magnetic pressure support for the vertical structure, the toroidal field reversal at $z\approx 0$ seen globally means that our magnetically elevated disks can remain gas pressure dominated in the disk midplane. This physical configuration for the toroidal field can occur in local simulations, but is not guaranteed to be present.

%====================================================
\begin{figure}
\centering
\includegraphics[width=0.96\columnwidth]{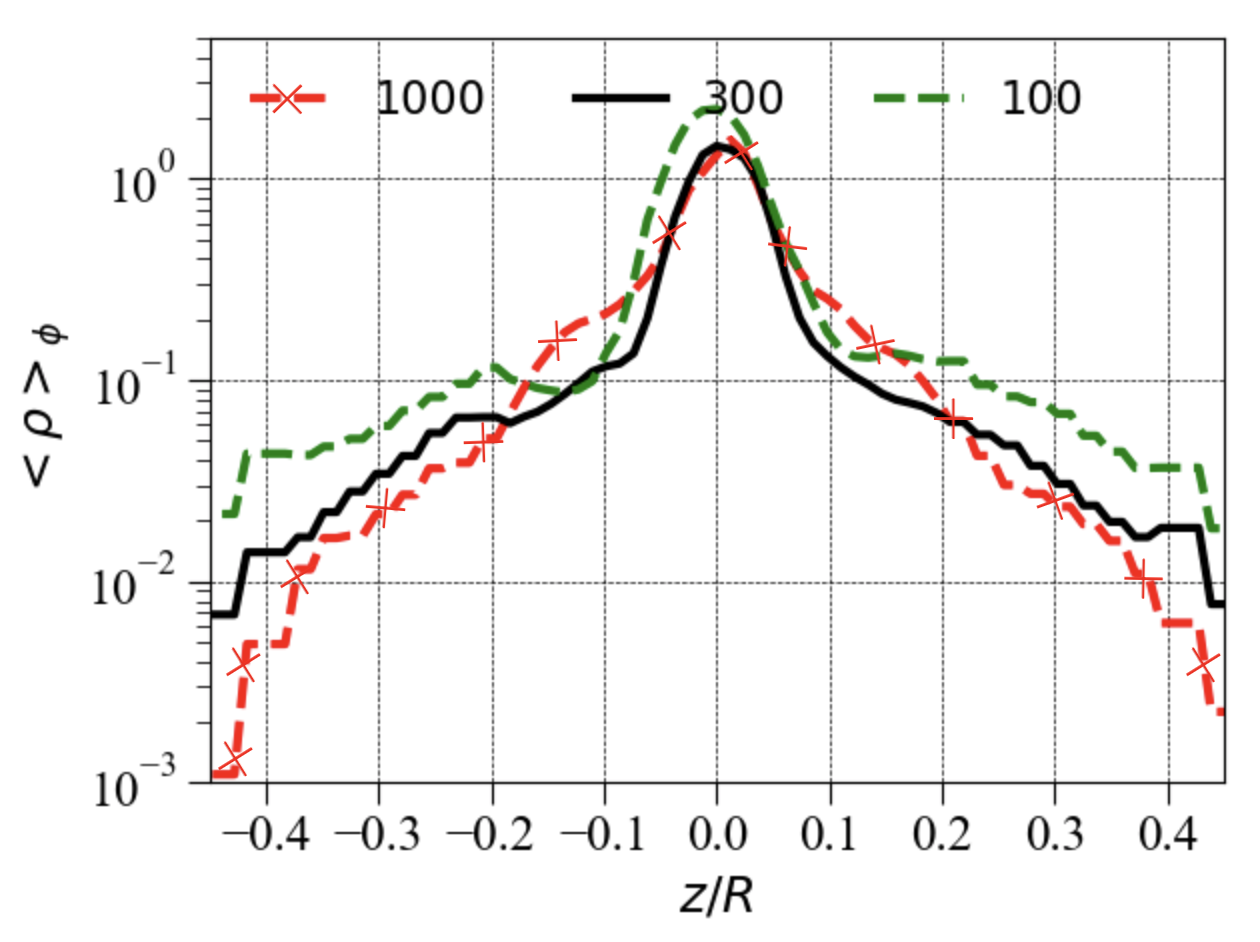}
\includegraphics[width=0.96\columnwidth]{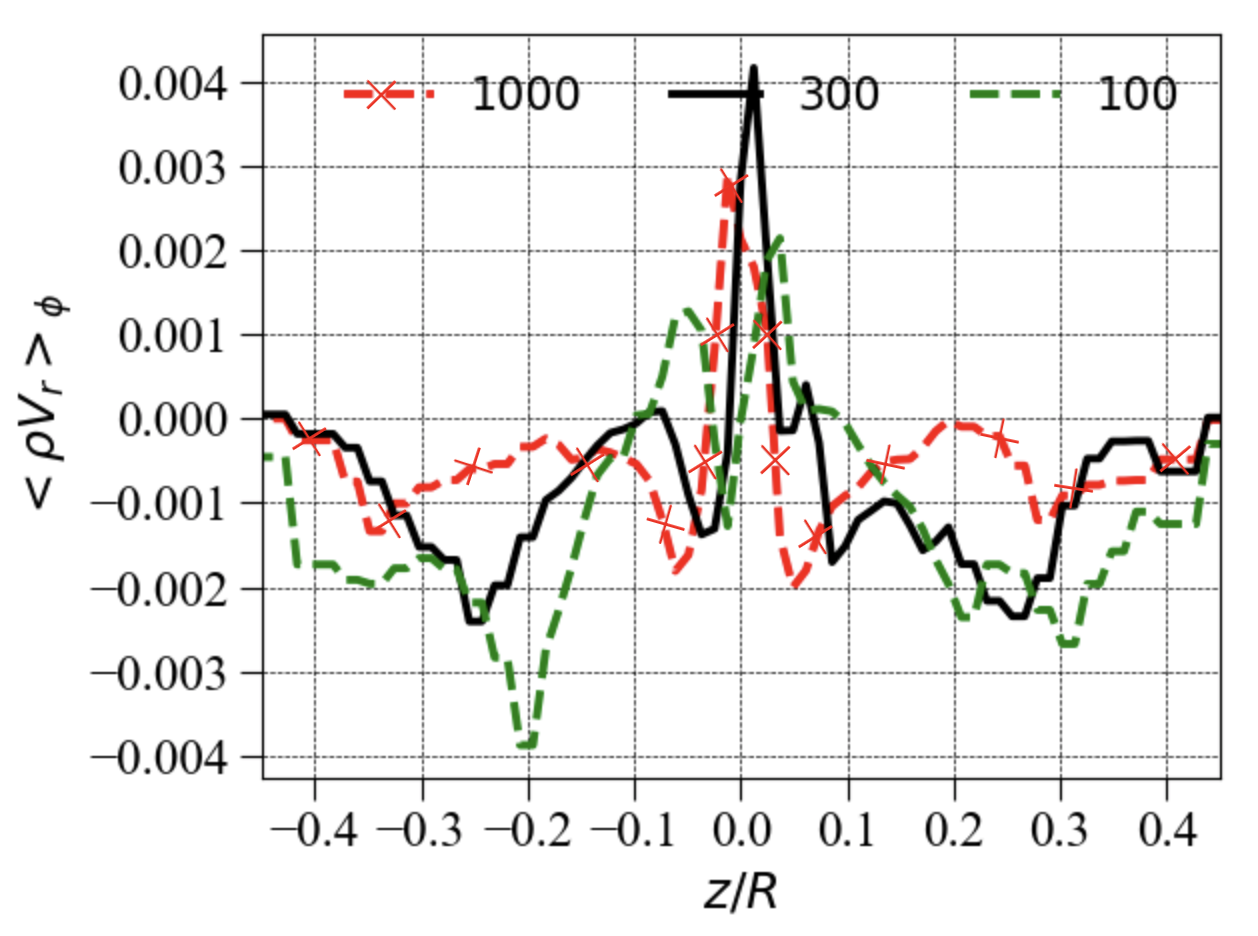}
\includegraphics[width=0.96\columnwidth]{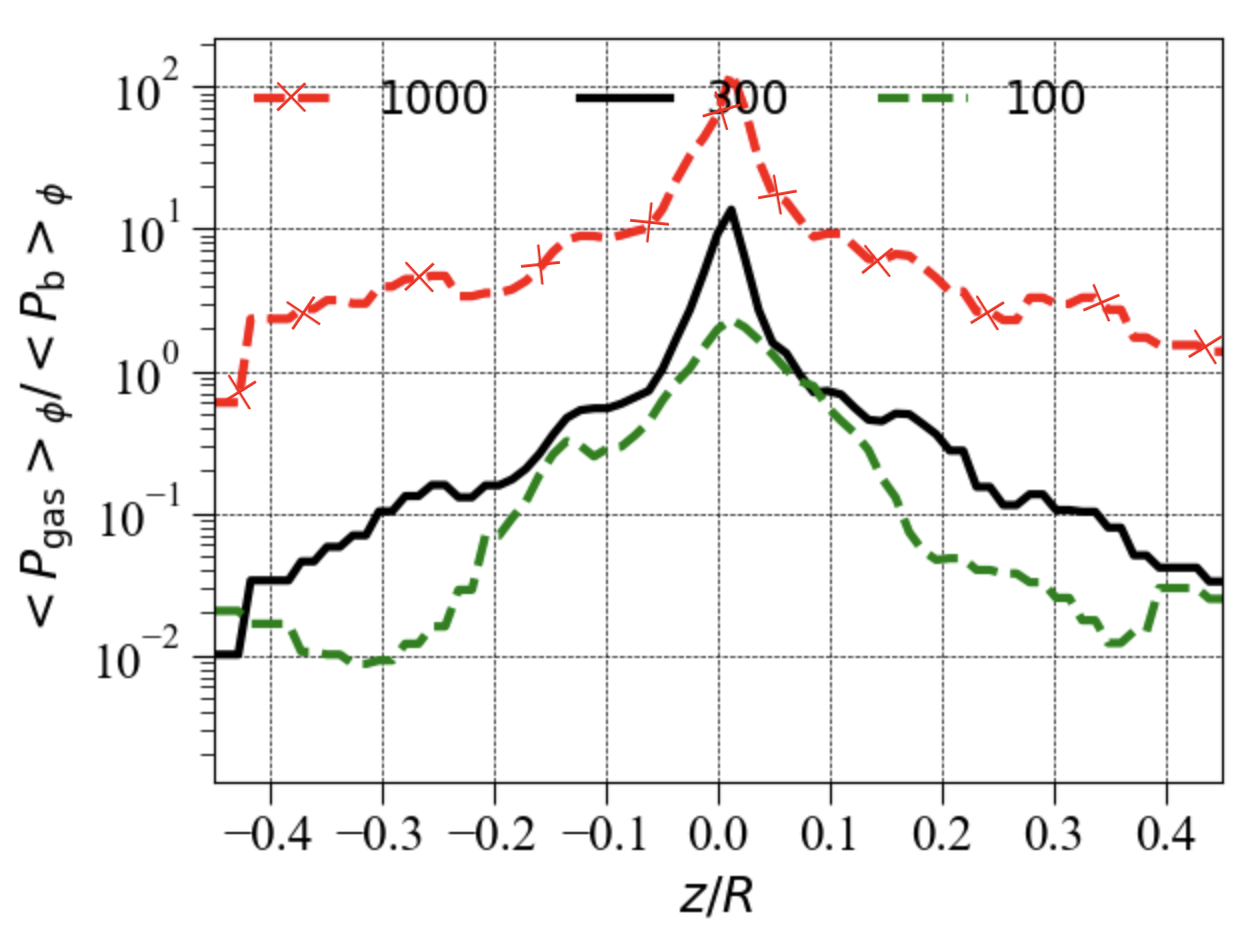}
\caption{Vertical profile of azimuthally and time averaged $\rho$, $\rho v_r$ and {\it total} $\beta$ ($\langle\beta_t\rangle = \langle P_\mathrm{gas}\rangle_\phi/\langle P_\mathrm{b}\rangle_\phi$) at $R = 1$, for different values of $\beta_0$. $\langle \beta_t\rangle$ is dominated by the toroidal component of the magnetic field. The red, black and green curves correspond to weak, intermediate and strong field cases, respectively. The time averaging is done from $t = 19$ to $t = 23$ orbits.}
\label{fig:betarhovr}
\end{figure}

%====================================================
\begin{figure}
\centering
\includegraphics[width=\columnwidth]{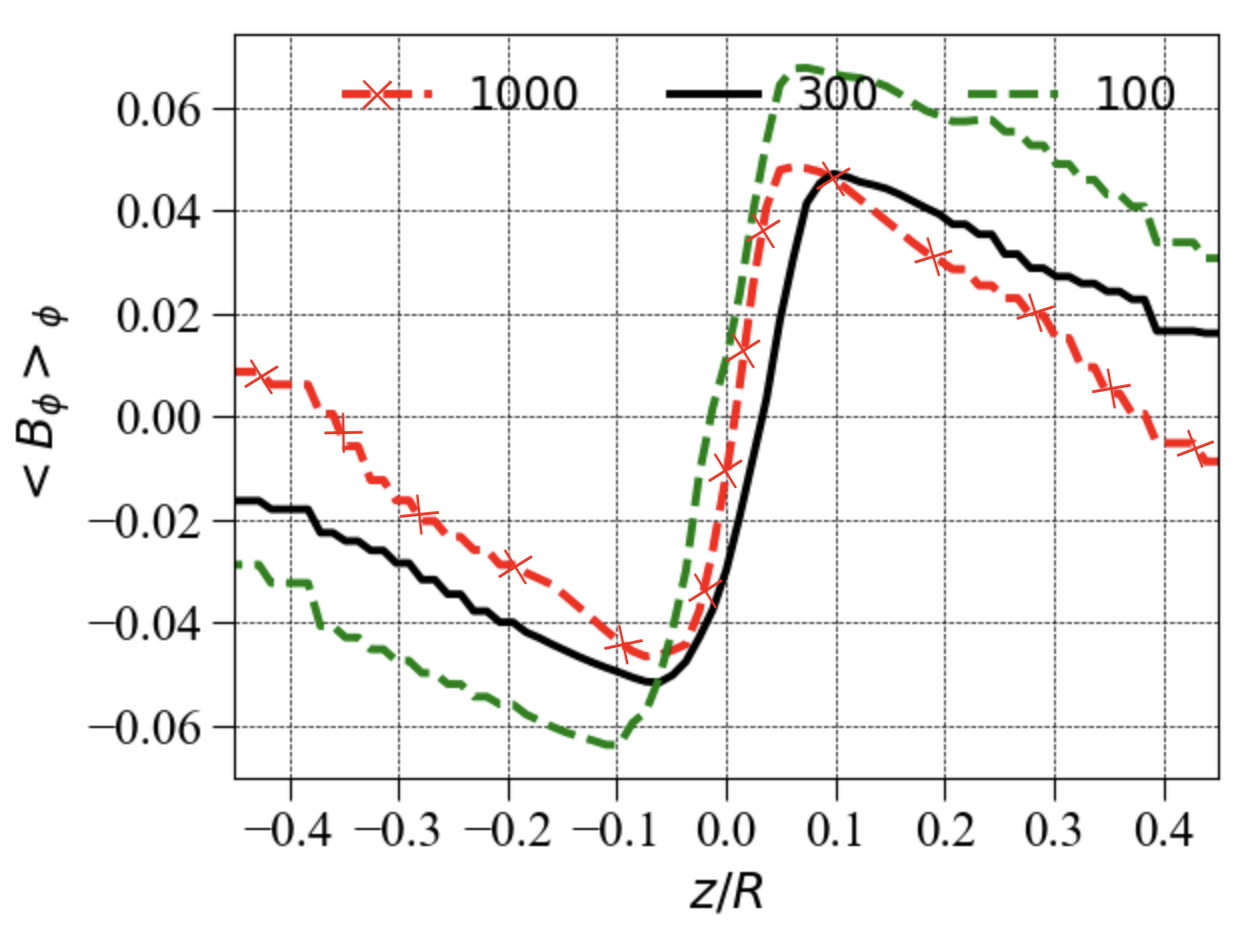}
\includegraphics[width=\columnwidth]{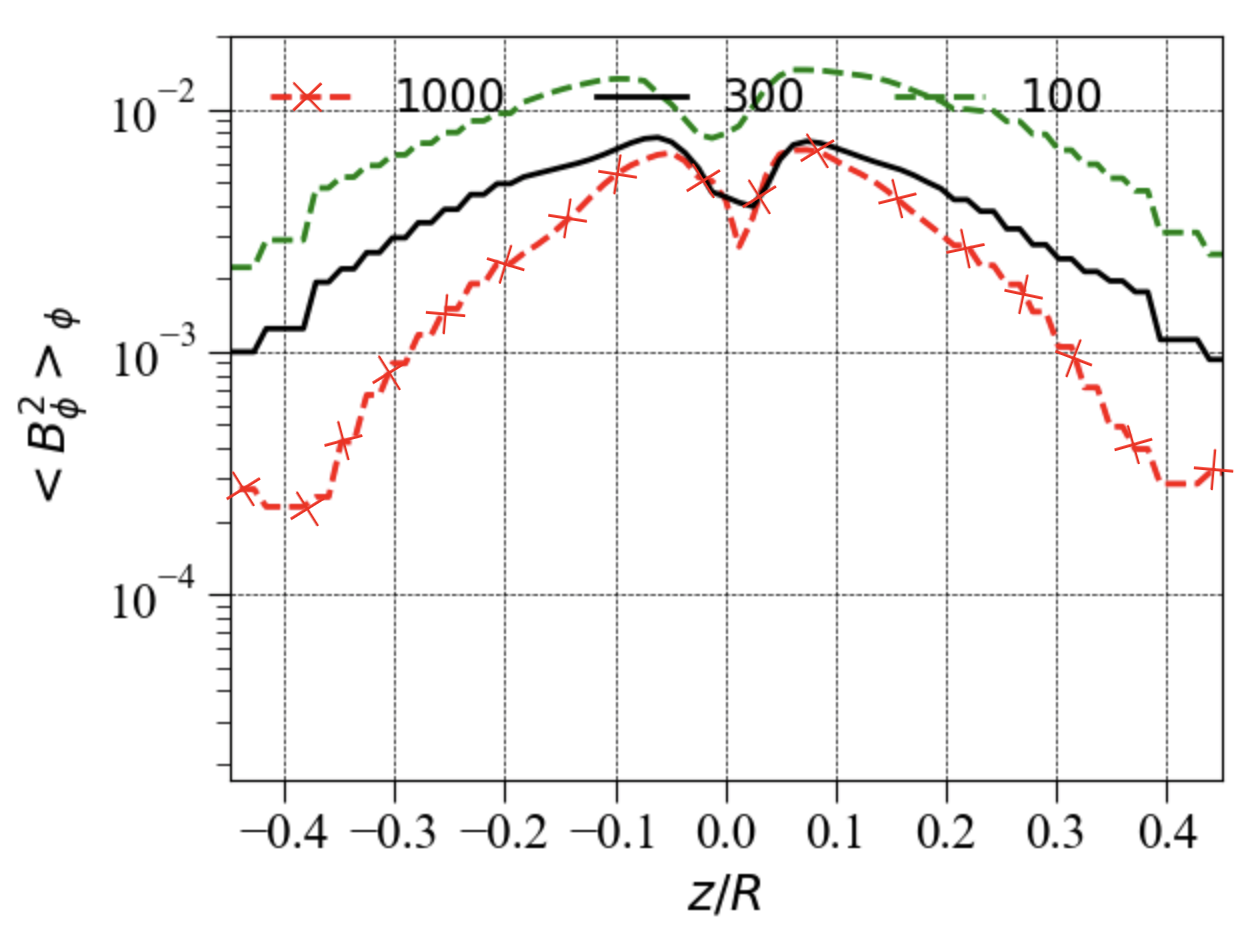}
\caption{Similar to Fig.~\ref{fig:betarhovr}, vertical profile of azimuthally and time averaged azimuthal component of the magnetic field, $B_\phi$, and $B^2_\phi$. The azimuthal and time averaging is done at local radius $R =1$. The time averaging is done from $t = 19$ to $t = 23$ orbital periods.}
\label{fig:bphi2}
\end{figure}

%=================================================
\begin{figure}
\centering
\includegraphics[width=0.98\columnwidth]{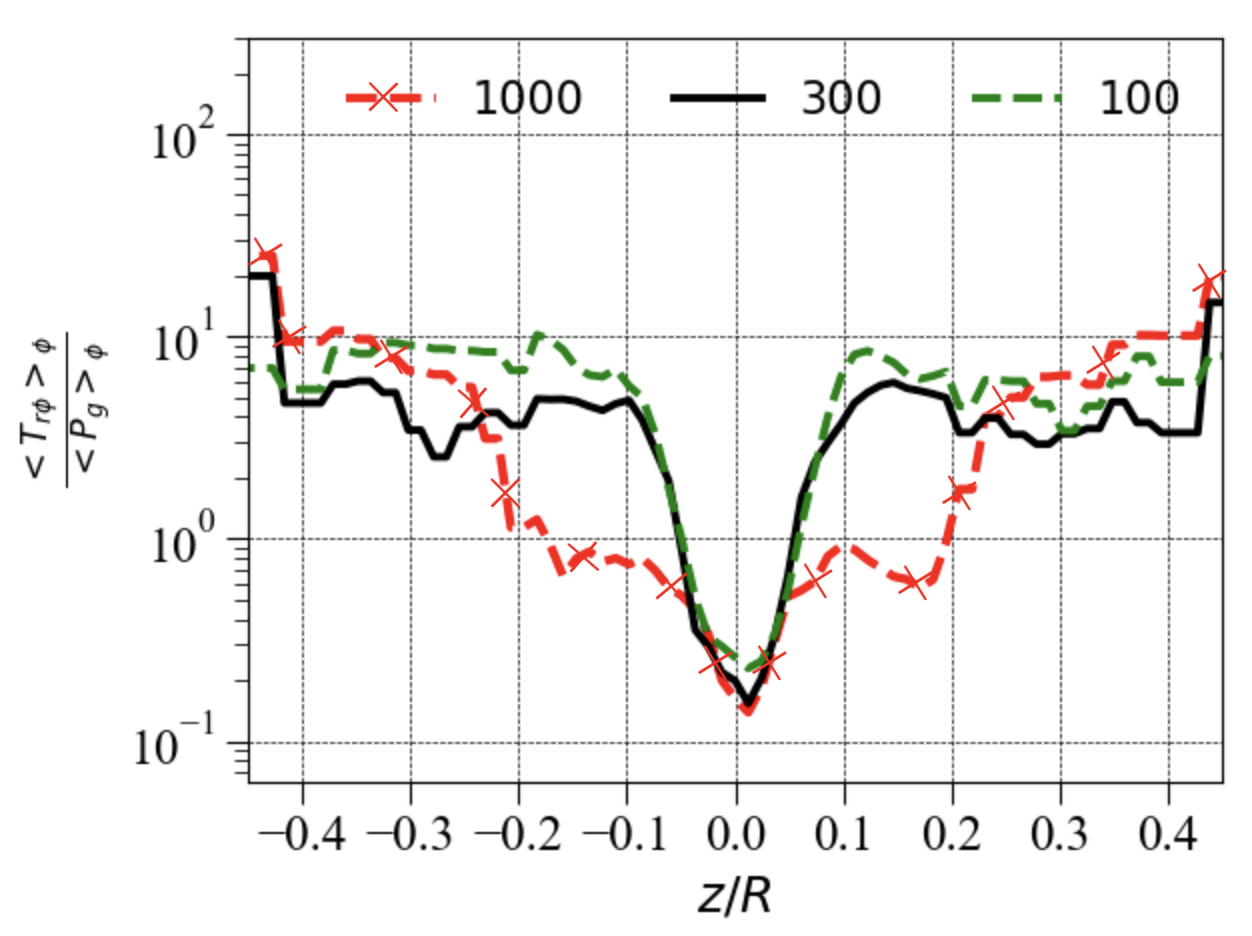}
\includegraphics[width=0.98\columnwidth]{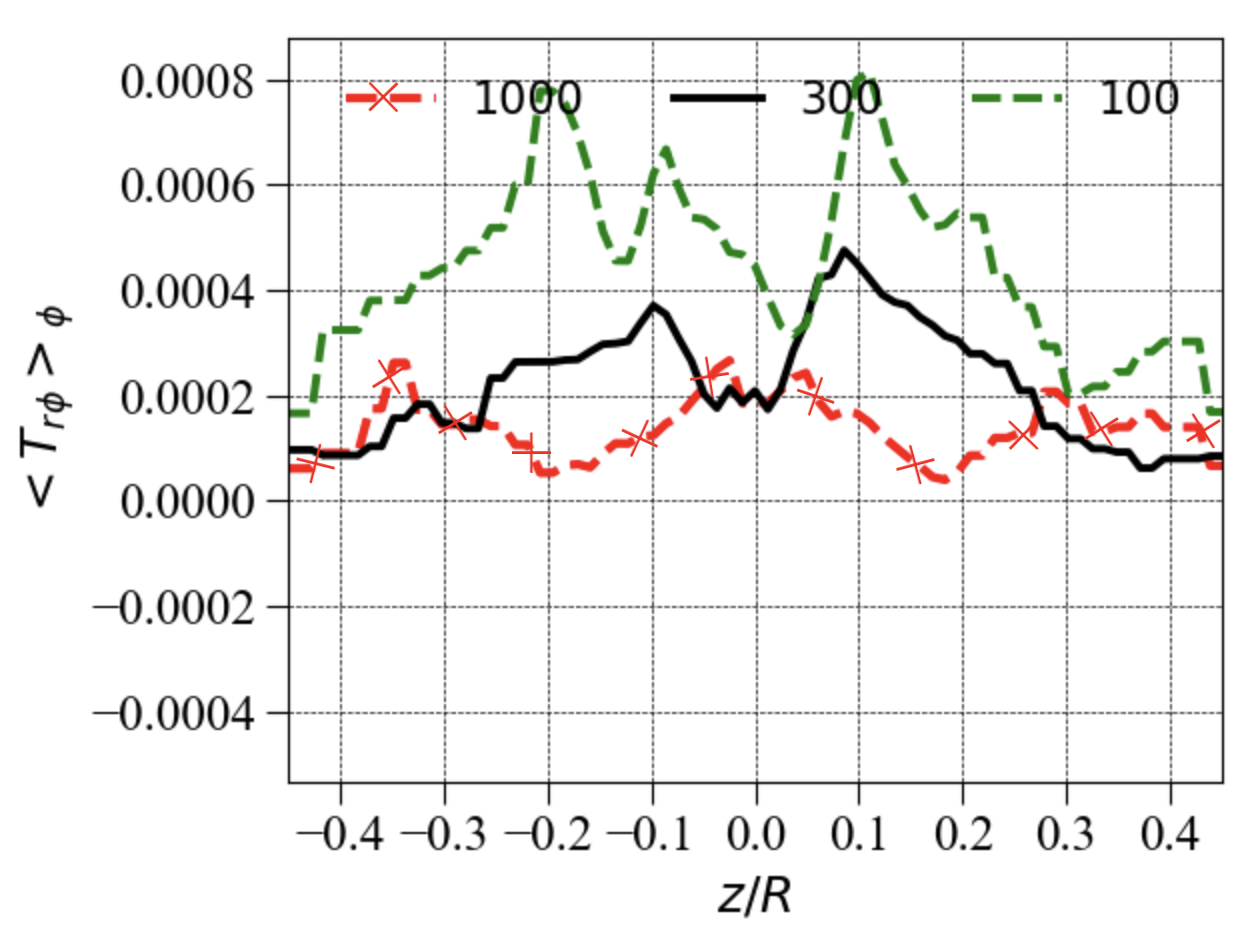}
\includegraphics[width=0.98\columnwidth]{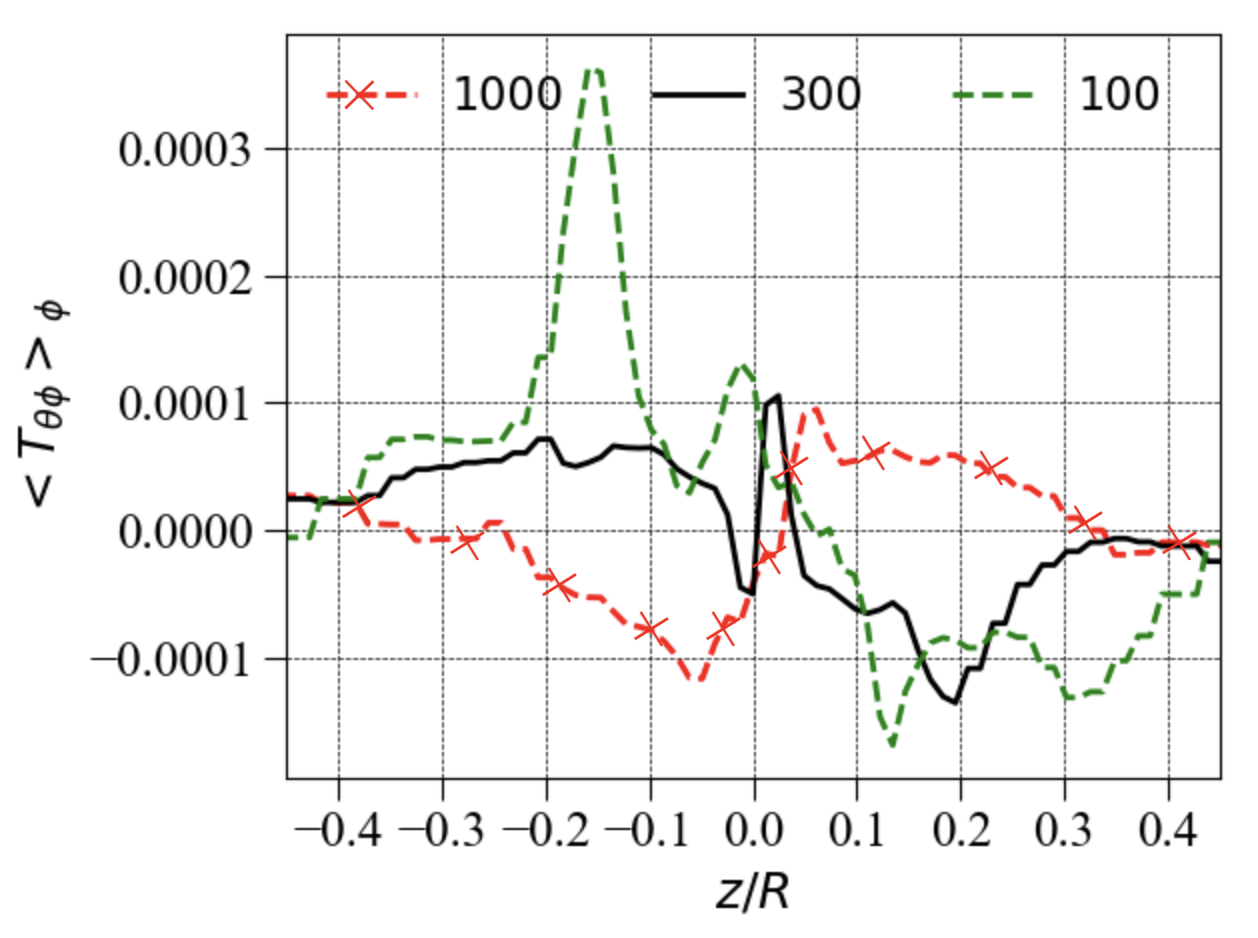}
\caption{Azimuthally and time averaged vertical profiles of viscosity parameter $\alpha = \langle T_{r\phi}\rangle_\phi/\langle P_\mathrm{gas}\rangle_\phi$, total radial stress $T_{r\phi}$ (sum of Reynolds and Maxwell stresses) and total vertical stress $T_{\theta\phi}$, which is again the sum of Reynolds and Maxwell stress at $R =1$. The time averaging is done from $t = 19$ to $t = 23$ orbital periods.}
\label{fig:stress}
\end{figure}
%=================================================
\subsubsection{Profiles of the disk variables}
Fig.~\ref{fig:betarhovr} shows the vertical structure of the simulated disks. The top panel shows the vertical profile of the density, which has a maximum at the disk midplane and a steeply decreasing profile up to $z/R ~\approx 0.1$. The density drops more gradually (compared to a Gaussian profile) at higher altitudes of the disk.
The density is highest in the elevated region for the strong field case and lowest in the weak field case. The results clearly indicate that the extent to which a disk will be magnetically elevated depends on the initial magnetic field strength.

The middle panel of Fig.~\ref{fig:betarhovr} shows the vertical profile of the radial mass flux density computed by azimuthally and time averaging (over orbits 19--23 for each magnetic field strength case) the product of the density and the radial component of velocity. The inflowing region appears at higher altitudes ($z/r \approx 0.2$, see Fig.~\ref{fig:streamline}), while the midplane shows outflow. The geometrically thin disk  ($H/r = 0.05$) case of \citet{ZhuStone} showed similar outflowing regions close to the midplane. The strong field case shows the largest accretion mass flux and the weak field case the smallest accretion flux close to $z/r \approx 0.2$. Unlike the strong field case and intermediate field case, the weak field model shows dominant accretion at $z/R \approx 0.1$ with a secondary smaller inflow region at $z/R \approx 0.3$.

The lowest panel of Fig.~\ref{fig:betarhovr} shows the vertical profile of azimuthally and time averaged total $\beta_t$, obtained by taking the ratio of the  azimuthal averages of gas and magnetic pressure at $R =1$. The disk is initialized with an initial $\beta_0 = 1000,\,300$ and $100$, and saturates with a final disk midplane total $\beta_t = 100,\,10$ and $2$ for the weak, intermediate and strong case, respectively. At higher altitudes in the disk, where accretion is preferentially occurring, magnetic pressure dominates in all three cases. In the accreting regions ($H/r \approx 0.2$), $\beta_t$ for the weak field case is approximately 4. The intermediate case achieves a final $\beta_t \approx 0.2$ at this height, and the strong field case attains a final $\beta_t \approx 0.08$. We find that the vertical profiles of total $\beta_t$ in our global simulations differ from those in the local shearing box simulations of \citet{Greg16}. The local shearing box simulations can achieve a much more strongly magnetized disk midplane because there is not necessarily an equatorial current sheet, in contrast to the global simulations (see Section~\ref{sec:localsim}). 

%====================================================
\begin{figure}
\centering
\includegraphics[width=\columnwidth]{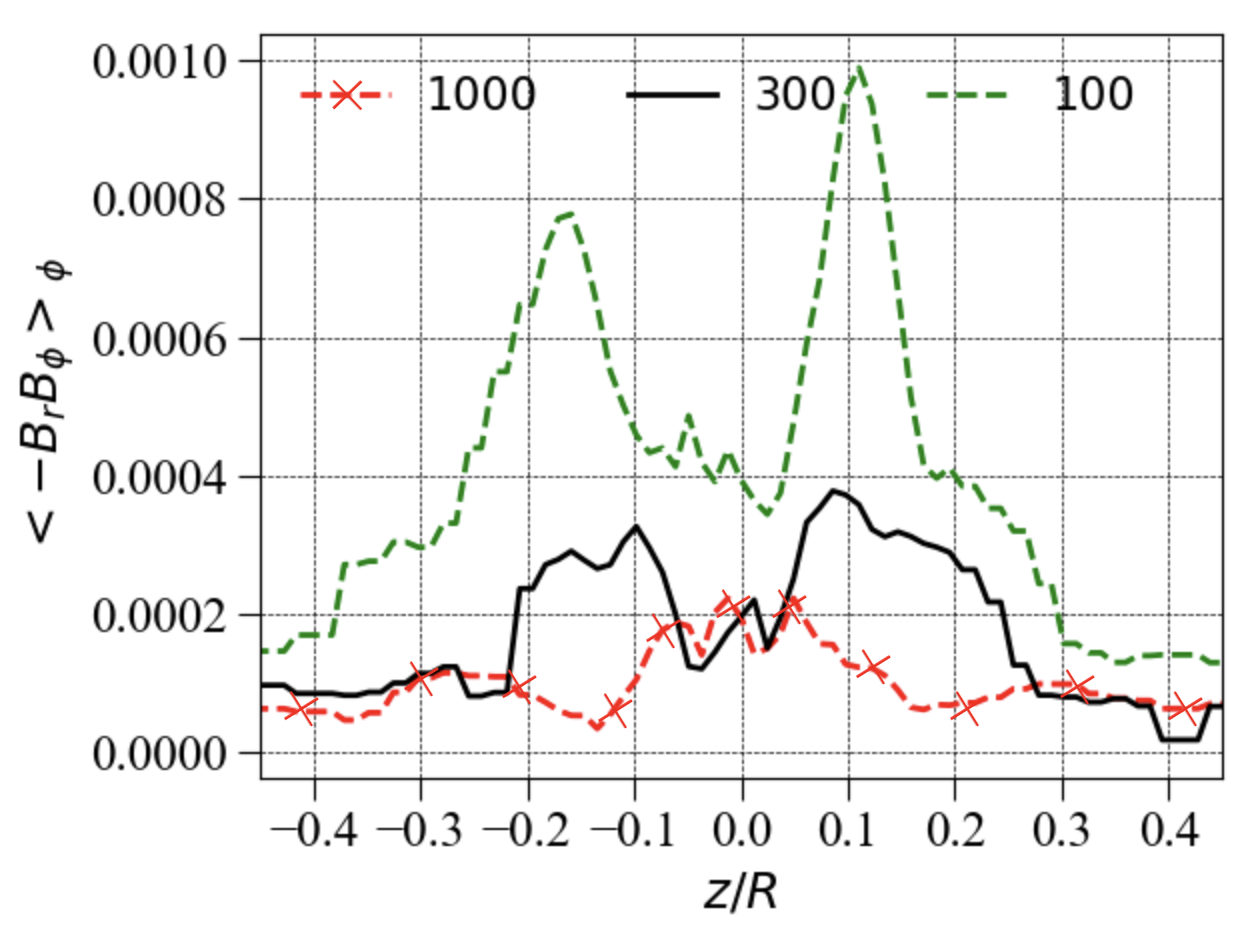}
\includegraphics[width=\columnwidth]{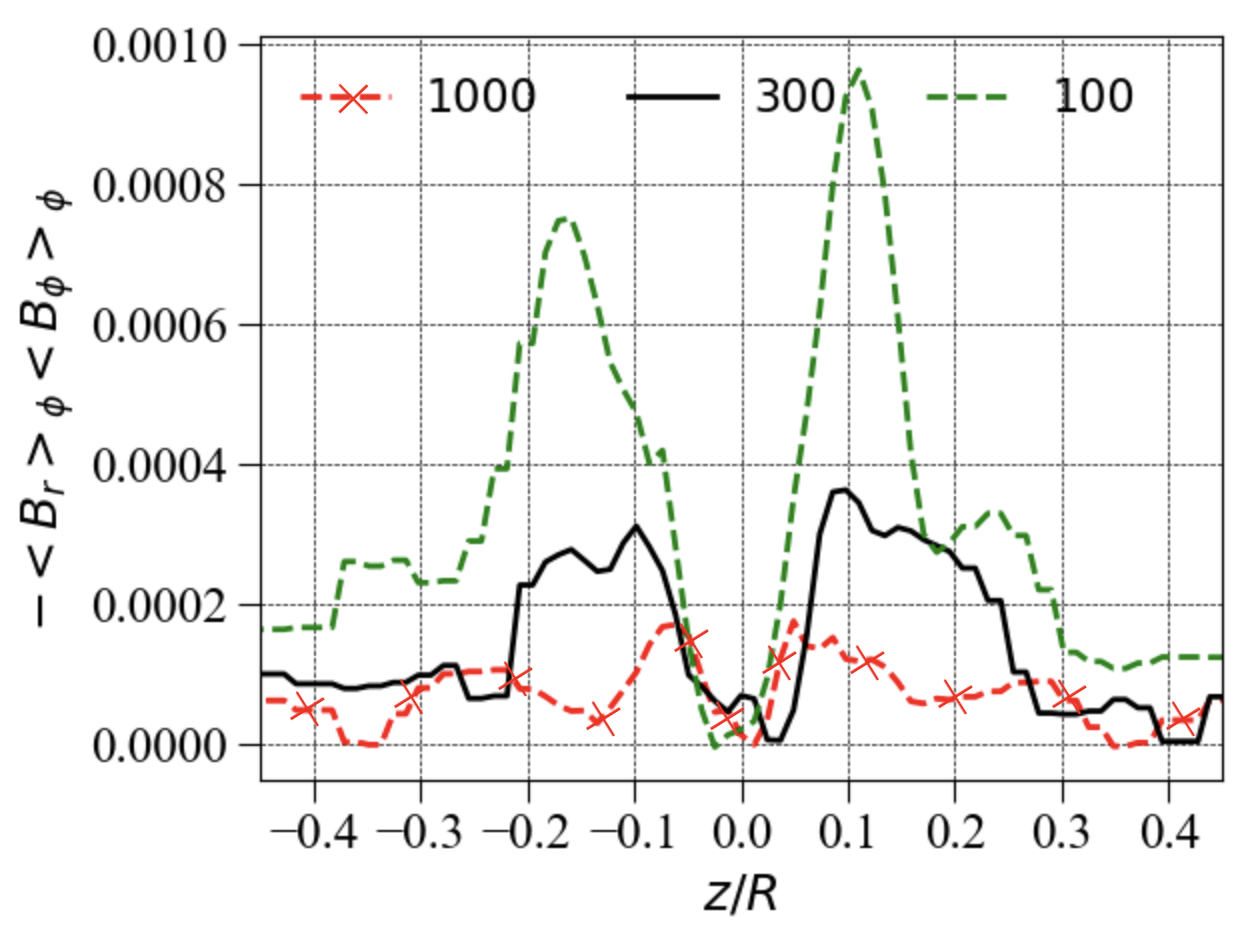}
\includegraphics[width=\columnwidth]{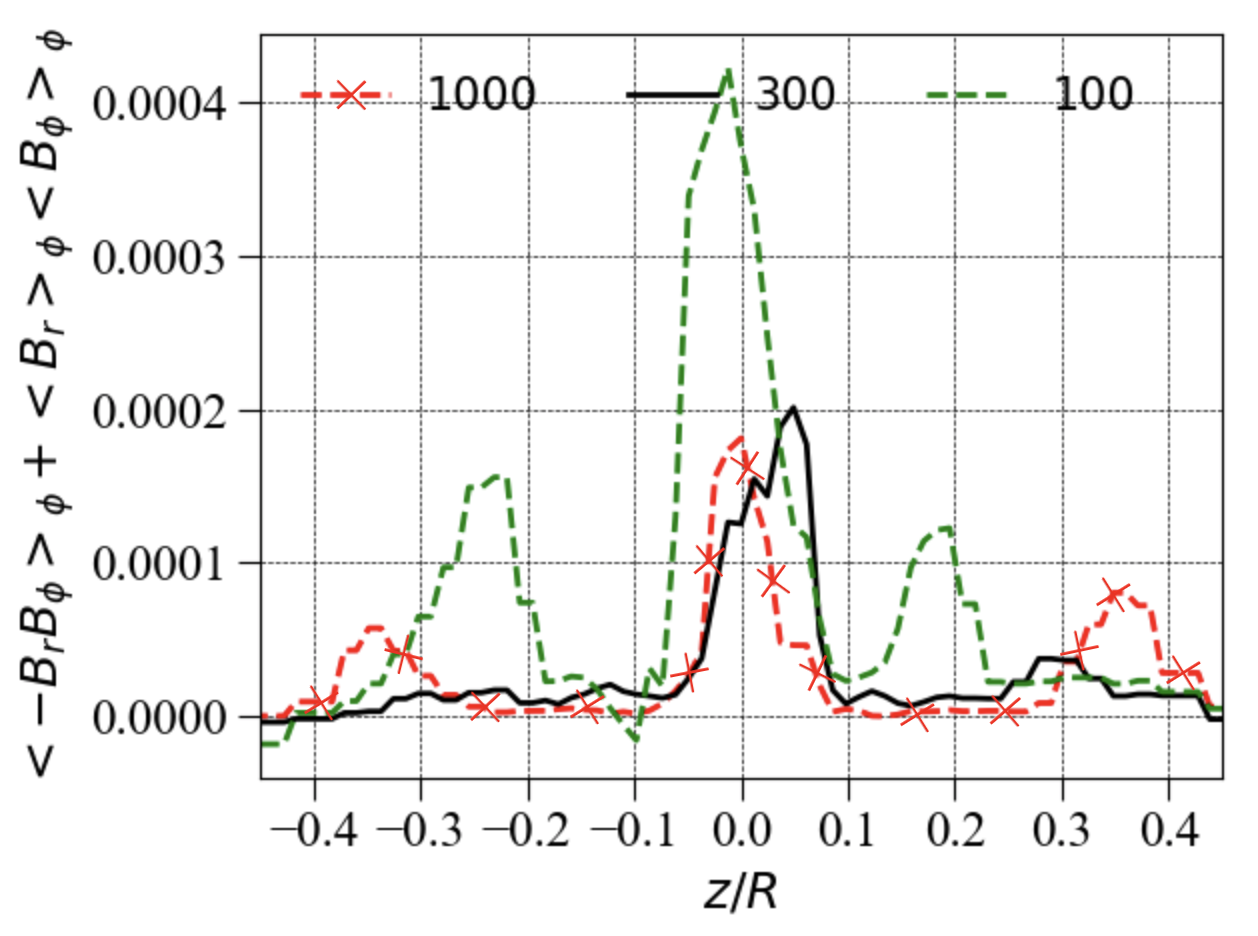}
\caption{Azimuthally and time averaged vertical profiles of total Maxwell stress (top panel), its coherent component (middle panel) and turbulent component (lowest panel) at $R =1$.}
\label{fig:stressM}
\end{figure}
%================================================
The azimuthally and time averaged vertical profile of $B_\phi$ (upper panel Fig.~\ref{fig:bphi2}) shows the dependence of the time-evolved toroidal magnetic field on the initial poloidal magnetic field strength. The weak magnetic field case ends up with a weak toroidal field and the strong magnetic field case leads to a stronger resultant toroidal magnetic field strength. Close to the disk midplane, $B_\phi$ is zero due to the formation of a current sheet there. This also gives intuitive and qualitative insight into how the toroidal field is being generated in these numerical models. Once the MRI triggers, an initial poloidal field develops a radial and toroidal magnetic field. MRI amplifies the toroidal field \citep{Latter10,Lessur13} and leads to a magnetic pressure dominated accreting region. In our chosen spherical polar coordinate system, a sheared radial and poloidal component of the magnetic field will produce a positive and a negative $B_\phi$ above and below the disk midplane, respectively. This is consistent with previous global simulations reported in \citet{ZhuStone}, but differs in significant respects from older models in which large-scale magnetic torques drove predominantly midplane accretion 
\citep{Pudritz85}. The lower panel of Fig.~\ref{fig:bphi2} shows $B^2_\phi$ to illustrate the gradual decrease of toroidal magnetic pressure with altitude. A dip in the vertical profile of $B^2_\phi$ close to the disk midplane is due to the current sheet. 
%==============================================
\subsubsection{Stress profiles in the disk}
 The time averaged viscosity parameter $\alpha$ is shown in the upper panel of Fig.~\ref{fig:stress} for each of our models. We emphasize that this value of $\alpha$ includes the total stress (sum of radial Reynolds and Maxwell stresses), in contrast to the standard $\alpha_{\rm SS}$ (Eqn.~\ref{eq:turbalpha}) which only includes the turbulent component. We find the expected trend that the weak initial magnetic field case gives a lower value of the viscosity parameter $\alpha$ and the strong magnetic field case gives larger values of $\alpha$. In the disk midplane, the difference in measured $\alpha$ from each of the simulations is small but in the {\it elevated} regions, where most of the accretion is occurring, the difference between inferred $\alpha$ between strong and weak field cases is much larger. The weak magnetic field case gives $\alpha \approx 1$ at $z \approx 0.2$, the intermediate initial magnetic field strength case gives $\alpha \approx 5$ and the strong magnetic field case gives $\alpha \approx 10$.  Note that these $\alpha$ parameters are normalized to the gas pressure, which is much lower than the magnetic pressure in the elevated layers for the intermediate and strongly magnetized cases. If the viscosity parameter is instead defined normalized to the magnetic pressure, all three cases have similar values in the accreting layers. 

In the middle panel of the Fig.~\ref{fig:stress} we show the time and azimuthally averaged total radial stress ($T_{r\phi} = T^\mathrm{Rey}_{r\phi} + T^\mathrm{Max}_{r\phi}$), which is dominated by Maxwell stress. The strong magnetic field case has almost a factor of eight larger value of total radial stress at $z/R \approx 0.2$ than the weak field case. The intermediate field strength case lies in between the two extreme cases of initial field strength. The polar stress ($T_{\theta\phi}$) is almost a factor of four smaller than the radial stress in the accreting regions. The vertical profile of the radial stress is always positive, implying a net outward angular momentum flux for Maxwell stress dominated flows. Above the disk midplane the azimuthal component of the magnetic field is positive and the radial component is negative (streamlines in Fig.~\ref{fig:streamline}) leading to positive radial stress above the disk midplane. Similarly, the vertical net flux seeding the magnetic field in our model leads to a positive radial component of the magnetic field (Fig.~\ref{fig:streamline}) and a negative azimuthal component below the midplane, also causing positive radial stress. This profile of radial stress ensures that there should be net accretion within the disk regions. 

The lowest panel in Fig.~\ref{fig:stress} shows the vertical profile of $T_{\theta\phi}$. A negative/positive value of $T_{\theta\phi}$ above/below the disk midplane implies an outward angular momentum carried away by winds. $T_{\theta\phi}$ behaves differently in the regions close to the disk midplane compared to higher altitudes. The green curve in this figure  shows large amplitude of negative and positive $T_{\theta\phi}$ above and below the disk region, respectively. The amplitude of $T_{\theta\phi}$ for higher altitudes in intermediate and weak field cases is much smaller compared to the strong field case. This indicates that the strong field case can launch winds more efficiently compared to the weak field case. Close to the disk midplane, $T_{\theta\phi}$ is pointing towards the disk midplane (specifically in the intermediate field case shown with solid black curve) and this could lead to angular momentum deposition on the disk midplane, causing outflows. These midplane outflows are not surprising because such backflows have also been predicted earlier in accretion disk solutions reported in \citet{KK02}. 
%=============================================

The total radial and polar stress is dominated by Maxwell stress. The total Maxwell stress can be decomposed into two components, one coherent (Eq.~\ref{eq:coherentM}) and the other turbulent (Eq.~\ref{eq:turbulentM}). The streamlines of the magnetic fields in Fig.~\ref{fig:streamline} show a coherent field configuration in elevated accreting regions. In order to quantify it further, we plot the total Maxwell stress, its turbulent component and coherent component in Fig.~\ref{fig:stressM}. Comparing the middle panel of Fig.~\ref{fig:stress} with the topmost panel of Fig.~\ref{fig:stressM} it is evident that the total radial stress is dominated by the Maxwell stress. Further decomposition of the Maxwell stress shows that it is dominated by the coherent component (shown in the middle panel of Fig.~\ref{fig:stressM}) everywhere except close to the midplane. Close to the disk midplane the coherent component vanishes. On the other hand, the turbulent component shown in the lowest panel dominates in the disk midplane but is about a factor of four smaller compared to the coherent component in the accreting regions. The turbulent $\alpha_\mathrm{SS}$ can be easily calculated given the total $\alpha$ in Fig.~\ref{fig:stress} and the turbulent Maxwell stress component. We find that $\alpha$ due to the coherent component dominates over turbulent $\alpha_\mathrm{SS}$. Our strong field and intermediate field cases are thus in qualitative agreement with \citet{ZhuStone}. The weak field case is quantitatively in agreement with \citet{ZhuStone}, where the coherent component of the stress is a factor of two larger than the turbulent stress in accreting regions.

%===================================================
\begin{figure}
\centering
\includegraphics[width=0.9\columnwidth]{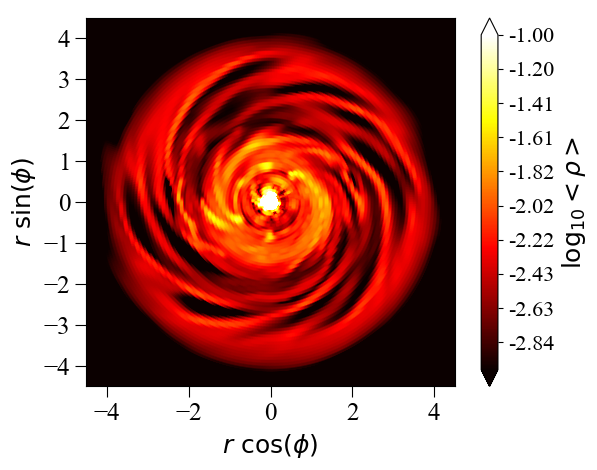}
\includegraphics[width=0.9\columnwidth]{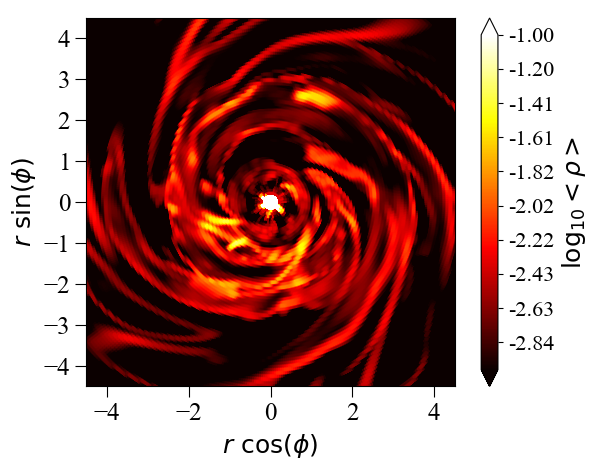}
\includegraphics[width=0.9\columnwidth]{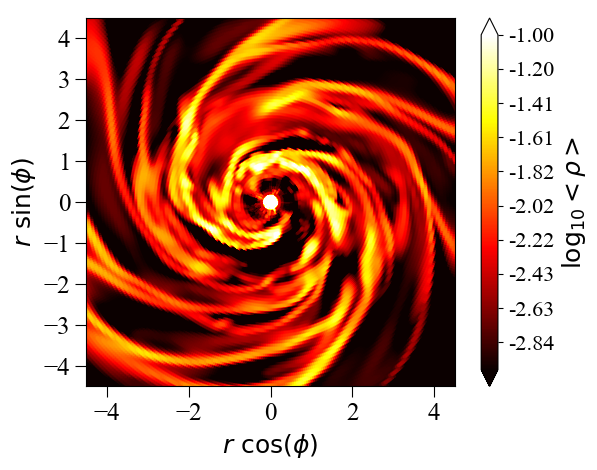}
\caption{Slices of mass density for $z = 0.2$ at $t = 20$ orbits (at $R=1$) for three initial $\beta$ models (top to bottom: $\beta_0 = 1000,\,300$ and $100$). The color-bar shows the logarithm of the density. The X- and Y- axes are labeled by cylindrical radius.}
\label{fig:dens_eq}
\end{figure}
%===========================================================================================
\begin{figure}
\centering
\includegraphics[width=\columnwidth]{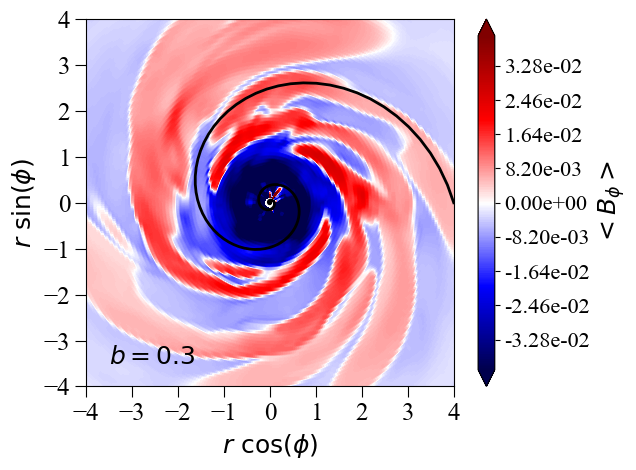}
\includegraphics[width=\columnwidth]{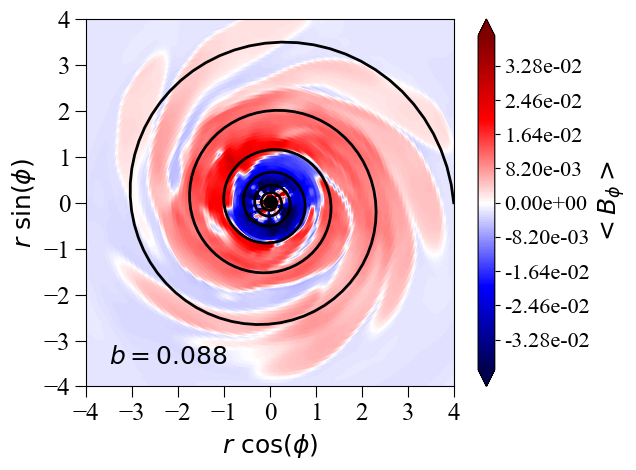}
\caption{Plots of the toroidal component of magnetic field for the intermediate ($\beta_0 = 100$, upper  panel) and strong ($\beta_0 = 300$, lower panel) magnetic field cases at $t = 20$ and $z = 0.2$. The solid black curves show logarithmic fits to the spiral structures seen in these two cases. The values for $b$ correspond to the winding parameters of these spirals. A larger value of $b$ means a more open spiral.}
\label{fig:spirals}
\end{figure}
%==========================================================================================
\subsection{Radial Structure}
\label{sec:radstructure}
In Fig.~\ref{fig:dens_eq}, we show off-equatorial ($z/R = 0.2$) slices of the mass density plotted at $t = 20$ orbital periods (measured at $R = 1$). The upper panel (initial $\beta_0 = 1000$) shows a weak spiral azimuthal density profile. The middle panel and lower panels show the density profile for initial $\beta_0 = 300$ and $\beta_0 = 100$, respectively. As we increase the initial magnetic field strength, the disk shows dramatically stronger spiral structures with an increasingly inhomogeneous density profile.
%===========================================================================================
\begin{figure}
\centering
\includegraphics[width=\columnwidth]{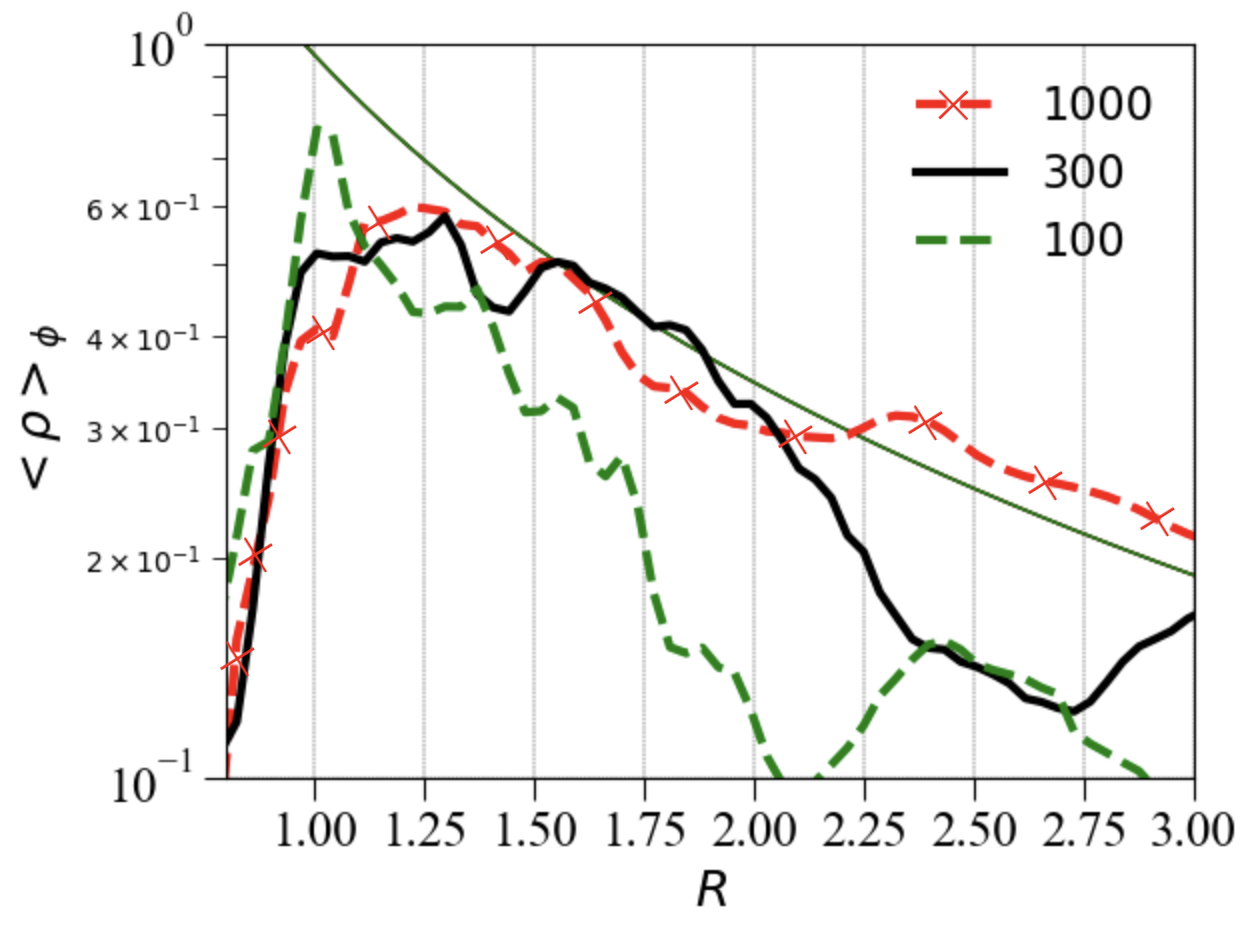}
\includegraphics[width=\columnwidth]{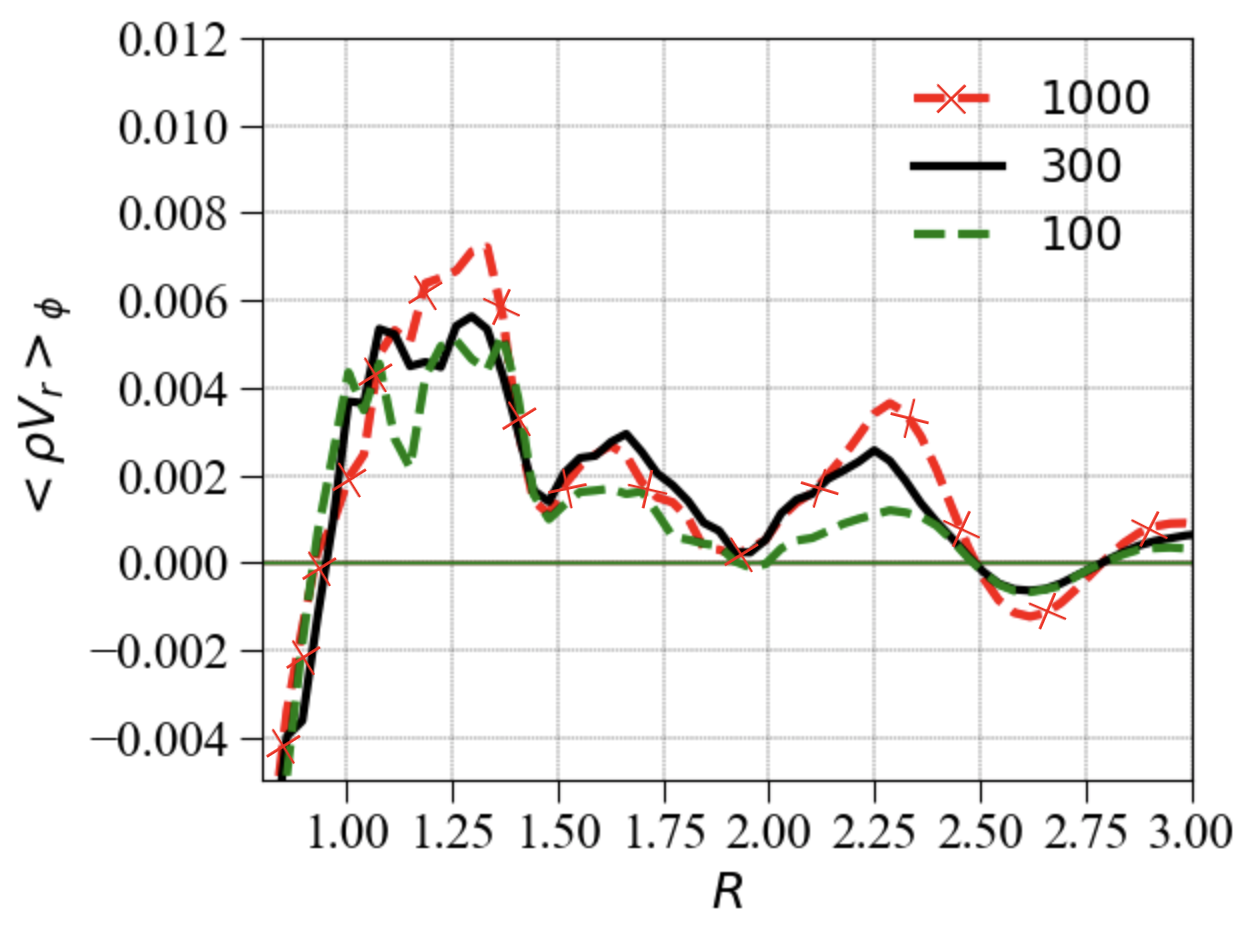}
\caption{Azimuthally and time averaged midplane radial profiles of density and mass inflow rate. Different curves show weak ($\beta_0 = 1000$), intermediate ($\beta_0 = 300$) and strong magnetic field ($\beta_0 = 100$) cases. The time averaging is done in the same way as vertical profiles for $t = 19$ to $t = 23$ orbits. The thin solid curve in the top panel corresponds to the initial midplane density profile of the disk. Positive values in the lower panel correspond to outflows and negative values to inflows.}
\label{fig:radialrho}
\end{figure}
%===========================================================================================
\begin{figure}
\centering
\includegraphics[width=\columnwidth]{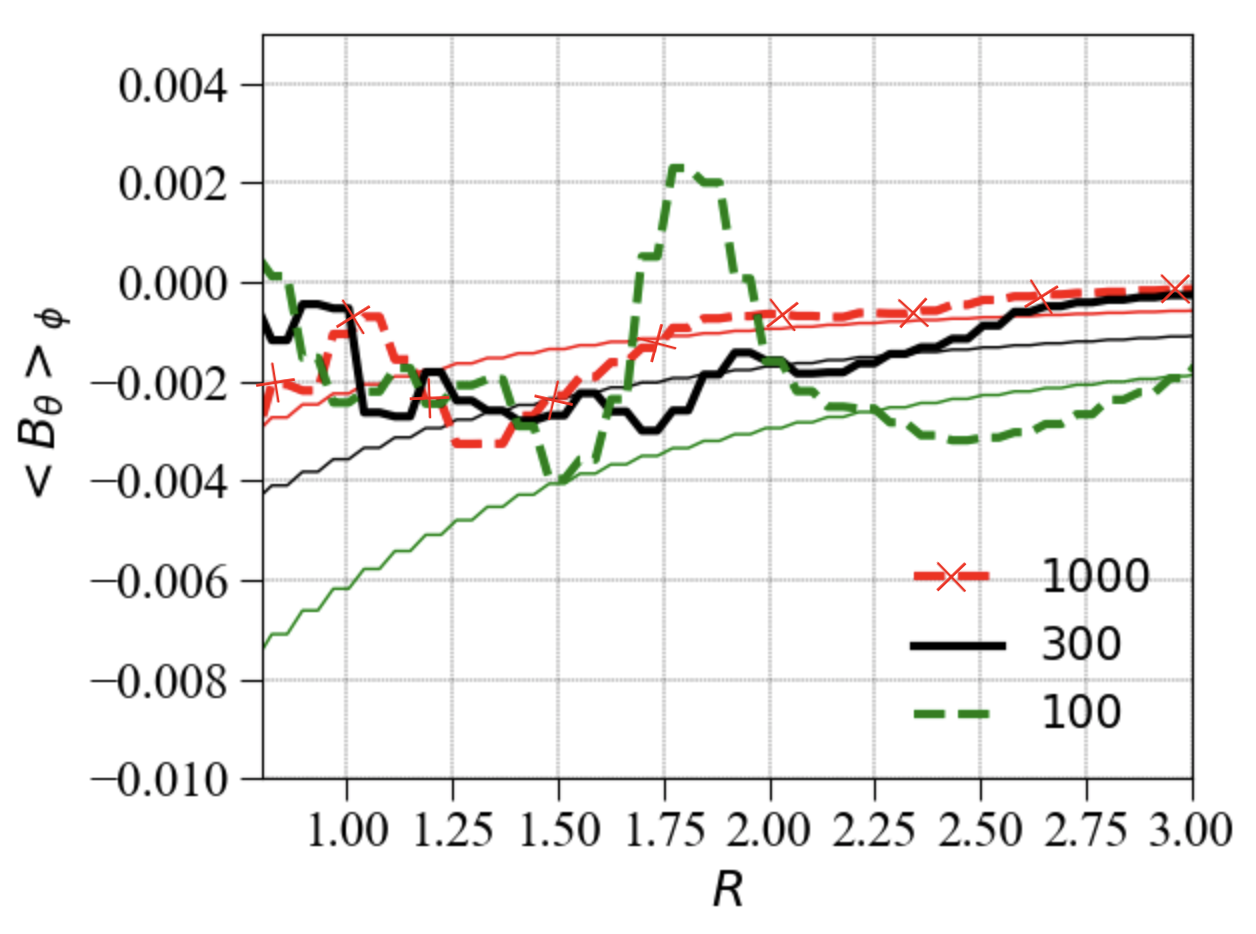}
\includegraphics[width=\columnwidth]{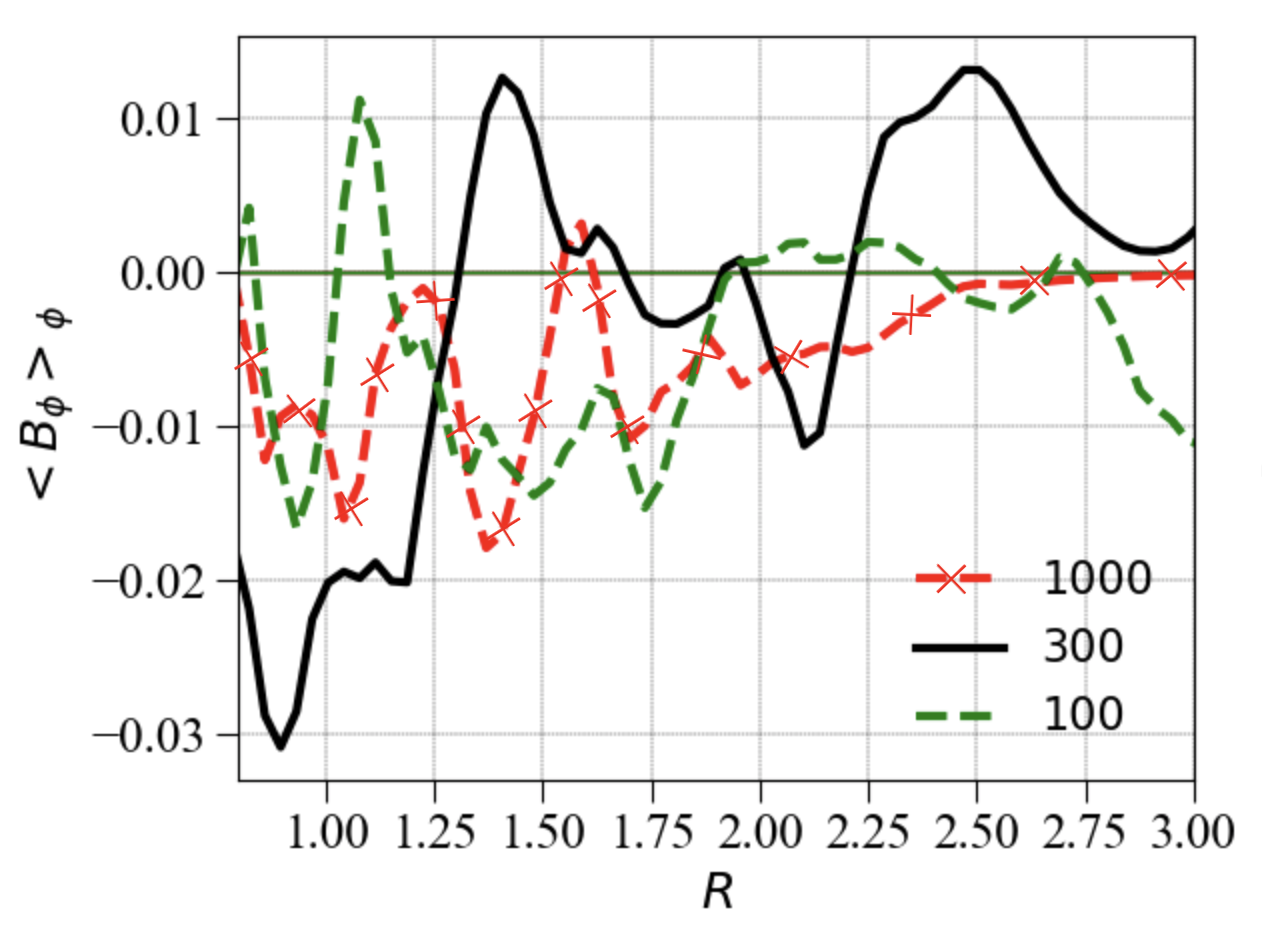}
\caption{Similar to Fig.~\ref{fig:radialrho} but for poloidal and toroidal components of the magnetic field. The thin solid curves in the upper panel correspond to the initial magnetic field radial profile for strong, intermediate and weak field case with green, black and red colors, respectively.}
\label{fig:radialb}
\end{figure}

%===========================================================================================
\begin{figure}
\centering
\includegraphics[width=\columnwidth]{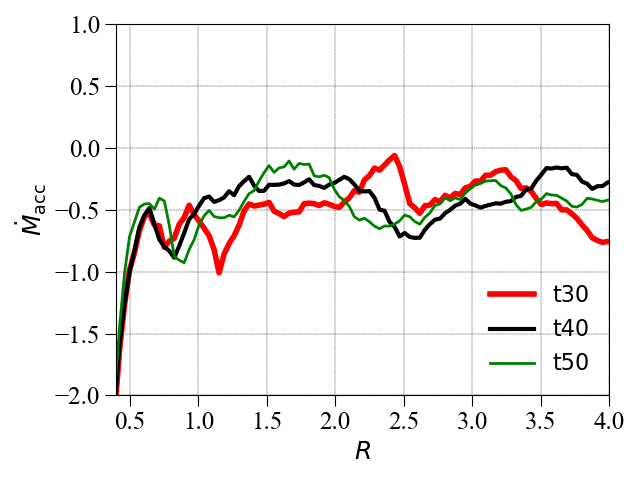}
\includegraphics[width=\columnwidth]{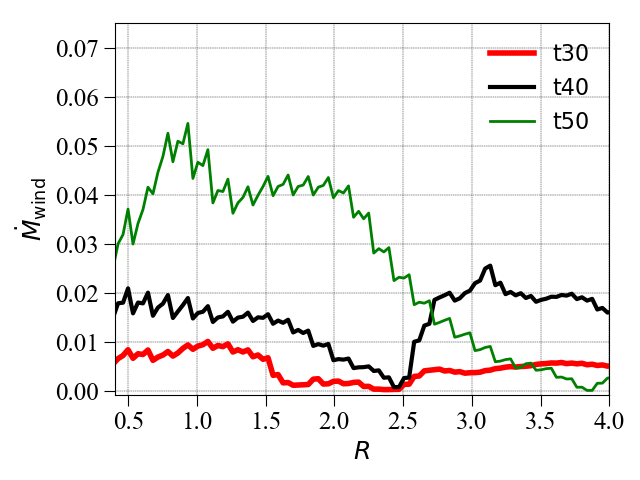}
\caption{Vertically integrated and azimuthally averaged radial profile of mass accretion rate (top) and mass flux due to winds (lower panel) for the strong field case in code units (density and velocity in units of $\rho_0$ and $v_K$ at $R =1$). Different curves show mass flux at $t = 30,\,40$ and $50$ orbital periods measured at $R =1$. The mass accretion rate is computed by integrating over the entire angular domain. Wind outflows are computed by integrating in the upper hemisphere for $\theta = 0-70^\circ$ and multiplying by 2 to account for winds from lower hemisphere.}
\label{fig:massflux}
\end{figure}
%=======================================
\begin{figure}
\centering
\includegraphics[width=\columnwidth]{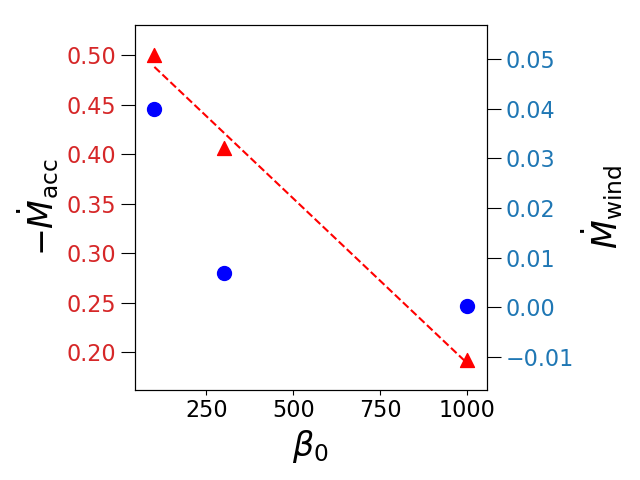}
\caption{Time averaged ($t=19-23$ orbits) mass accretion rate and mass loss rate at $r=1$ as a function of initial magnetic field strength ($\beta_0$). Note the different scales for mass accretion rate and wind mass loss rate. The red triangles show mass accretion rate and blue circles mass loss rate. The horizontal axis is shared for both left (red) and right (blue) vertical axis for mass accretion rate and mass loss rate respectively. The dashed red line shows best fit for mass accretion rate.}
\label{fig:mdot_beta}
\end{figure}
%===========================================================================================
\begin{figure}
\centering
\includegraphics[width=\columnwidth]{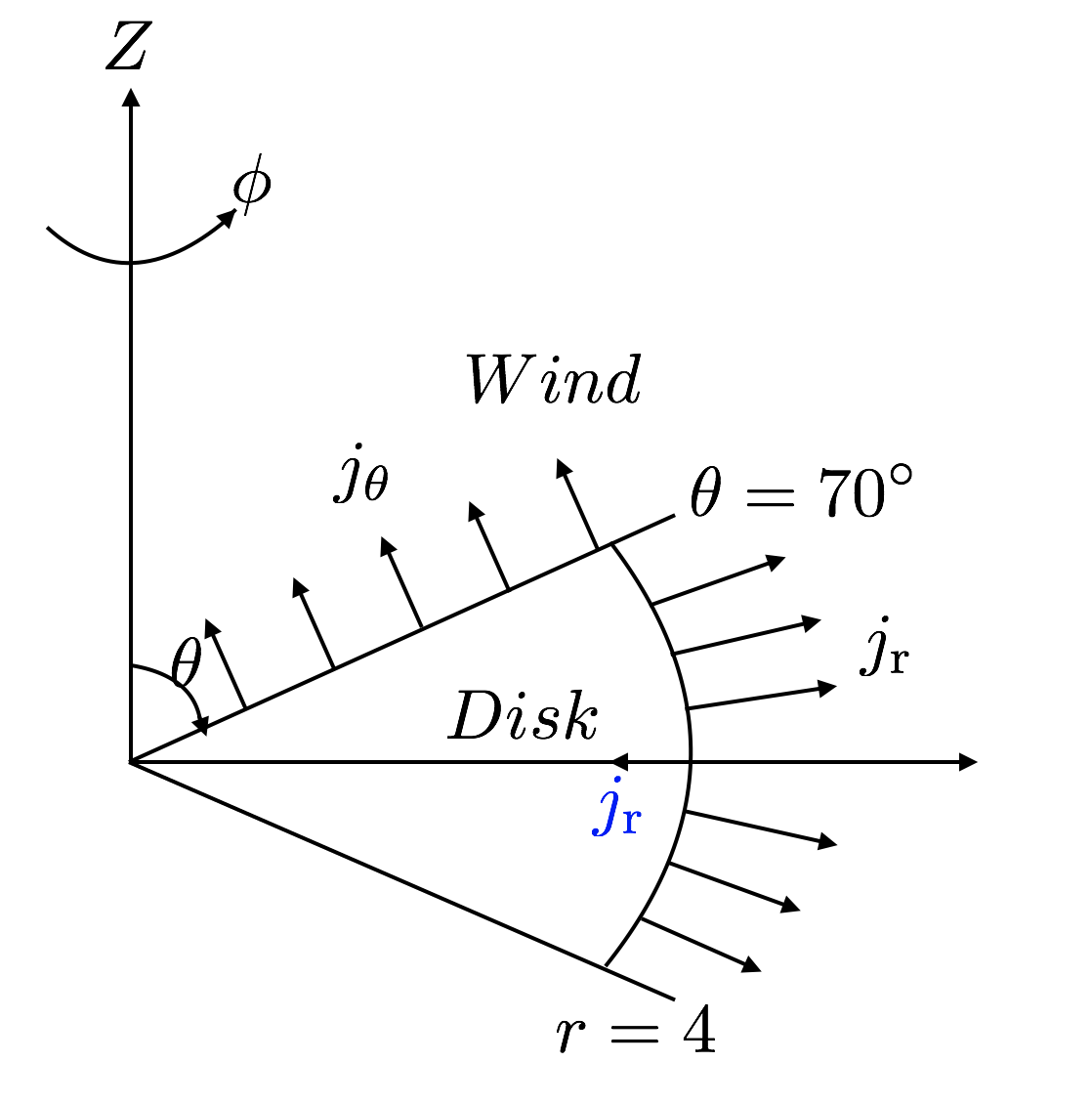}
\includegraphics[width=\columnwidth]{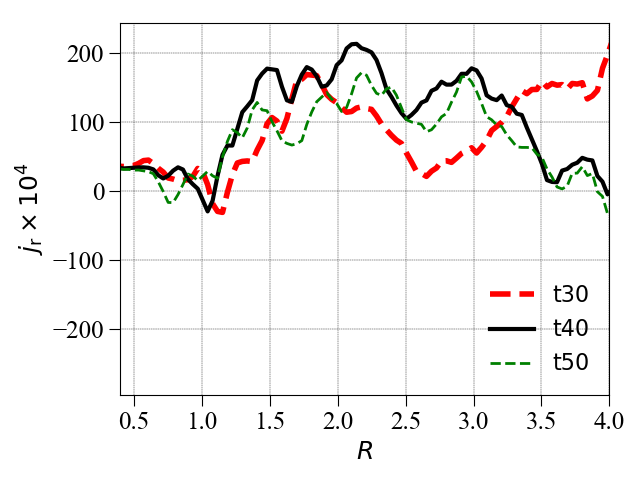}
\includegraphics[width=\columnwidth]{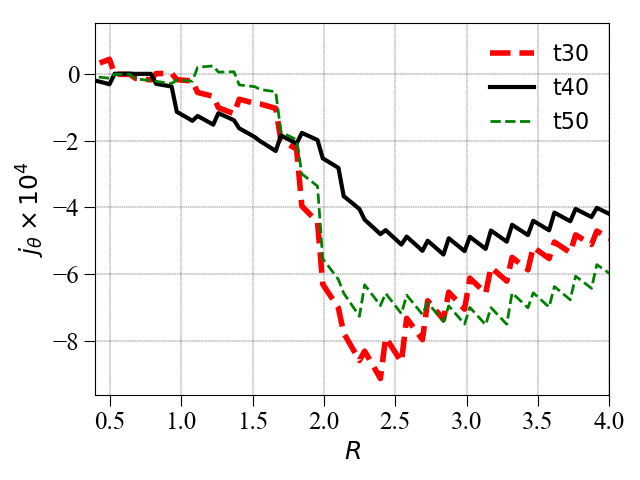}
\caption{Angular momentum flux due to winds and accretion in units of Keplerian value at $R =1$. The top panel shows a schematic diagram of the two surfaces we chose to compute the angular momentum flux. The positive $j_\mathrm{r}$ in the middle panel implies that angular momentum is being transported outward in the accretion dominated disk. Negative $j_\theta$ in the lower panel, on the other hand, shows angular momentum is being carried away in winds. (Note that the unit vector along $\theta$ points toward the disk and negative flux is therefore directed away from the disk.) The vertical axis shows angular momentum flux multiplied by $10^4$ to avoid decimals. A unit value of angular momentum in the vertical axis will be equivalent to 0.0001 in code units.}
\label{fig:jflux}
\end{figure}
%====================================================
\begin{figure}
\centering
\includegraphics[width=\columnwidth]{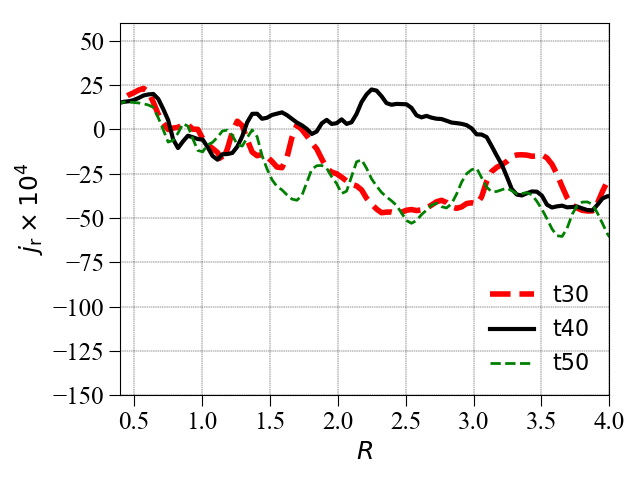}
\caption{Angular momentum flux close to the disk midplane. The flux is computed using Eq.~\ref{eq:momr} but integrated only within 8 cells in $\theta$ close to the disk midplane. The unit of angular momentum flux is the same as in Fig.~\ref{fig:jflux}.}
\label{fig:midplaneout}
\end{figure}
%====================================================
In Fig.~\ref{fig:spirals} we plot the toroidal component of the magnetic field at higher altitude ($z/R = 0.2$) where most of the accretion is occurring. We only show the toroidal field slice for the strong and intermediate field cases. The weak field case shows a corresponding density profile but does not show significant inhomogenities. The color-bar shows the range of the toroidal magnetic field using (blue, red) for (negative, positive) $B_\phi$.

The strong ($\beta_0 = 100$) and intermediate ($\beta_0 = 300$) magnetic field cases show spiral structures that can be approximately fitted with logarithmic spirals of the functional form $r = a e^{- b\phi}$, where $a$ and $b$ are constant coefficients for a chosen setup. In Fig.~\ref{fig:spirals}, the black curves show the fits to the spiral arms in the azimuthal component of the magnetic field, with $b = 0.3$ in the upper panel ($\beta_0 = 100$) and $b=0.088$ in the lower panel ($\beta_0 = 300$).  This trend is consistent with $b \propto \beta_0^{-1}$. The fit is created by taking the logarithm of $B_\phi$ to eliminate the negative component and distinguish or isolate positive $B_\phi$ regions in Fig.~\ref{fig:spirals}. We trace the the ridge of the field strength and fit a logarithmic spiral through it.  A detailed study of the formation of these spirals will be reported in an upcoming paper.

The midplane radial profiles of the azimuthally averaged density, mass flux, poloidal field ($B_\theta$) and toroidal field ($B_\phi$) are shown in Figs.~\ref{fig:radialrho} and \ref{fig:radialb}, respectively. The weak and intermediate field cases have higher density at the disk midplane compared to the strong field case. Since the strong field case accretes at higher rate, the low midplane density simply reflects the fact that matter is draining more quickly. 

The lower panel of Fig.~\ref{fig:radialrho} shows the product of the density and radial component of velocity at the disk midplane. We notice both outflowing ($R>0.95$) as well as inflowing ($R<0.95$) structures. This is consistent with the vertical profile of the same variable (Fig.~\ref{fig:betarhovr}). Accretion is mainly occurring in the surface of the disk (see Fig.~\ref{fig:streamline}), with negligible accretion along the disk midplane (except very close to the inner boundary). The vertically integrated mass flux in Fig.~\ref{fig:massflux} also shows a dominant accretion across all the radii and a significantly large inflow inside ($R<0.5$), which is very close to outflowing inner boundaries.

In Fig.~\ref{fig:radialb}, we find an enhanced magnetic field strength in the inner regions. Note that we overplot the initial profile of the magnetic field strength  $B_\theta \propto R^{-5/4}$ (and $B_\phi = 0$) as thin curves. The poloidal component of the magnetic field has flipped its orientation at $R<1.0$ for each magnetic field case. The strong field case also reverses its sign at larger radii. The bottom panel shows that the toroidal field strength dominates over the poloidal component. In both poloidal and toroidal magnetic field radial profiles, we find sign reversals. This has also been seen in previous global simulations of weakly magnetized disks by \citet{ZhuStone}. As we discussed earlier in connection with the vertical stress profiles, the disk midplane is dominated by turbulent stress, leading to large fluctuations in magnetic field orientation.  

%================================================
\begin{figure}
\centering
\includegraphics[width=1\columnwidth]{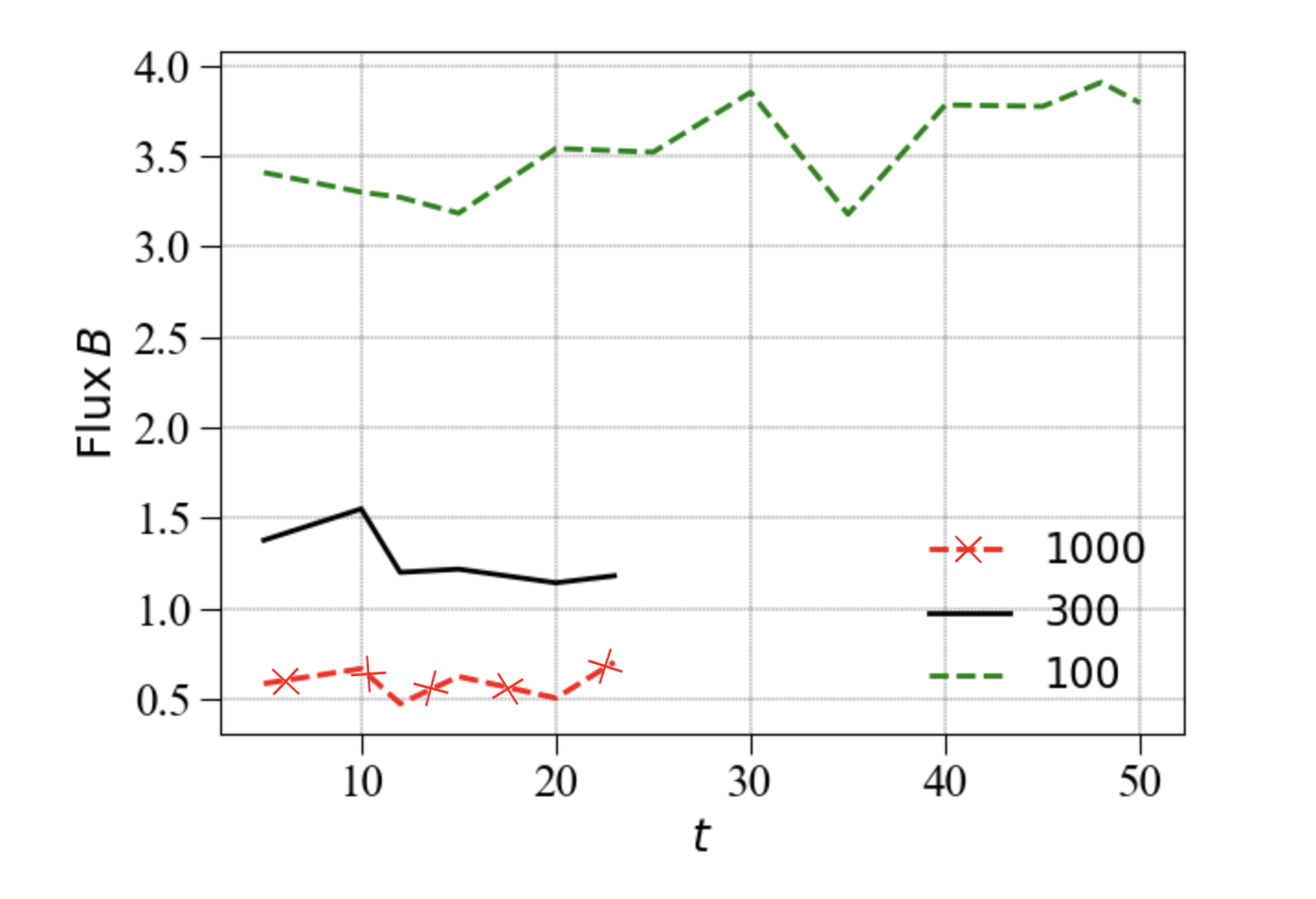}
\caption{The time evolution of the magnetic flux for weak, intermediate and strong field cases. The flux is computed within a cylinder $0.1< R < 1.0$ (excluding flux from the lower surface of the cylinder). The time on the  horizontal axis is measured in units of the orbital period at $R =1$.}
\label{fig:flux_evolution}
\end{figure}
%================================================
%================================================
\subsection{Accretion, winds and outflows}
\label{sec:inflowoutflow}
We find a dominant accretion flow in all three models we simulated. However, in addition to inflow, the disk also exhibits winds (in polar regions) and outflows (disk midplane). If we closely look into the flow above or below the disk midplane, it has three distinct regions as shown in the top panel of Fig.~\ref{fig:jflux}. The disk midplane shows outflows followed by elevated accreting regions at $z/R = 0.2$. Above these accreting regions, there are winds which are distinct from the midplane outflows. This suggests that the vertical structure of the disk is much more complicated than just turbulent inflowing gas in the disk and winds along the polar regions. In this subsection, we focus on each of these individual flows. First we discuss accretion, secondly winds and finally midplane outflows. 

In order to compute the mass accretion rate, we used the vertically integrated mass flux at each spherical radius,
\begin{equation}
    \label{eq:inflows}
    \dot{M}_\mathrm{acc}(r) = \int^{\pi}_0\int^{2\pi}_0 \rho v_r r^2 \sin\theta d\theta d\phi \ . 
\end{equation}
We evolved the strong field case for $50$ orbital periods at $R = 1$, which gives a steady state accretion flow up to $R \approx 3$. This can be seen in Fig.~\ref{fig:massflux}, where we show the radial profile of mass accretion rate for the strong field case. The three different curves in Fig.~\ref{fig:massflux} show an almost horizontal radial profile of the mass flux for three time windows separated by 10 orbital periods (measured at $R =1$) each. We also note that, despite large outflow velocity magnitudes at high latitudes shown in Fig.~\ref{fig:streamline}, the flow is dominated by accretion. A large inflow rate in the $R< 0.5$ regions is influenced by inner outflowing boundary conditions. This region is also influenced by density floor values applied to maintain a reasonable time step ($10^{-7}$) and to avoid magnetic flux loss near the poles. 

It is also important to estimate the mass and angular momentum carried away by winds from the accretion disk. In order to compute the mass outflow rate from the wind regions, we use a polar domain spanning from $\theta = 0$ to $\theta =70^\circ$.
\begin{equation}
    \dot{M}_\mathrm{wind}(r) = \int^{70^\circ}_0 \int^{2\pi}_0 2\rho v_r r^2 \sin\theta d\theta d\phi.
    \label{eq:wind}
\end{equation}
The factor of 2 in  Eq.~\ref{eq:wind} accounts for the wind mass loss rate from the lower hemisphere. The lower panel of Fig.~\ref{fig:massflux} shows that in our strong field case, mass loss due to the wind varies between $2-6\%$ of the total accreted mass at different time windows separated by 10 orbital periods at $R=1$. This also suggests that winds are not steady but time varying. A time averaged mass loss rate due to winds  (Fig.~\ref{fig:mdot_beta}) shows that about $6\%$ of the accreted mass is carried away by winds.

In computing the mass loss rate due to winds, we encounter low resolution regions close to the boundary of the third and second refinement levels, leading to small fluctuations in the radial profile of the wind mass loss rate. Despite this caveat due to low resolution in the disk wind regions, we find an order of magnitude larger wind driven mass loss rate compared to \citet{ZhuStone} for their initial $\beta = 1000$ case. This suggests that strongly magnetized accretion disks can launch winds more efficiently compared to weakly magnetized accretion disks. 

We further quantify this in Fig.~\ref{fig:mdot_beta} by plotting the mass accretion rate and mass loss rate due to winds as a function of initial $\beta_0$. The red triangle data points show mass accretion rate ($\dot{M}_\mathrm{acc}$) and blue data points show mass loss rate ($\dot{M}_\mathrm{wind}$) computed at $r =1$. Note the different scales for mass accretion and loss.  Both mass accretion and mass loss rates decrease as we decrease the magnetic field strength.

We also estimate the angular momentum budget in winds and the disk region. We use the following diagnostics, to compute the radial angular momentum flux
\begin{equation}
    j_\mathrm{r}(r) = \int^{110^\circ}_{70^\circ}\int^{2\pi}_0 \left( \rho v_\mathrm{r}  v_\phi R  -R B_\mathrm{r}B_\phi\right) r^2 \sin\theta d\theta d\phi
    \label{eq:momr}
\end{equation}
and the polar angular momentum flux
\begin{equation}
    j_\theta(r) = 2 \int^R_{0.1} \int^{2\pi}_0\left( \rho v_\theta v_\phi R - R B_\theta B_\phi \right) rdr \sin\theta d\phi.
    \label{eq:momth}
\end{equation}

Positive $j_\mathrm{r}$ corresponds to angular momentum flux pointing outwards whereas negative $j_\theta$ computed at fixed $\theta = 70^\circ$ corresponds to flux pointing away from the disk. Comparing radial and polar angular momentum fluxes, we find that approximately $7\%$ of the angular momentum is carried away by the winds. Unlike the wind region, the disk region is dominated by accretion with outflows at the disk midplane. We also compute the angular momentum flux from a domain very close to the disk midplane by using 8 cells in $\theta$. Similar to the schematic diagram, we find that the angular momentum flux is pointing inwards at the disk midplane with fluctuations across zero. Comparing the middle panel of Fig.~\ref{fig:jflux} (which shows the total outward pointing angular momentum flux between $\theta = 70^\circ$ and $\theta = 110^\circ$) and Fig.~\ref{fig:midplaneout}, we find positive radial flux in the former and negative in the latter. The angular momentum flux at the midplane is pointing inwards, carrying approximately 15\% of the total outward pointing angular momentum flux shown in Fig.~\ref{fig:jflux}. This is also qualitatively consistent with \citet{ZhuStone}, where the authors found that approximately $10\%$ of the angular momentum is carried away by outflows at the disk midplane.  At small radii, the midplane angular momentum flux points outwards, suggesting that the disk midplane is weakly accreting there. The black curve in Fig.~\ref{fig:midplaneout} fluctuates across zero values at the disk midplane, causing more complex inflowing and outflowing structures at the midplane.

%================================================
\subsection{Quasi-steady state of Magnetic Flux}
\label{sec:fluxBtime}
Given the large effect of net magnetic flux on disk structure, it is important to determine how the flux evolves with time as a function of disk radius \citep{Begelman14}. If the accretion flow advects magnetic field effectively, a large magnetic flux will accumulate close to the accreting object and eventually may lead to a magnetically arrested disk \citep{Narayan03,McKinney12}. Alternatively, rapid diffusion of the field will lead to weakly magnetized inner regions that are more prone to thermal instability (in X-ray binaries) and disk fragmentation in the AGN context. Here we address the competition between advection and diffusion that we observe in our global simulations. Our simulations include the poles and hence we do not lose magnetic flux from polar boundaries. This makes these simulations ideal to study magnetic flux evolution in strongly magnetized disks.

In Fig.~\ref{fig:flux_evolution} we show how the poloidal magnetic flux evolves in an enclosed cylinder with $0.1 < R < 1.0$. The green dashed curve shows the strong field case which we have evolved for $50$ orbits. The magnetic flux in this case varies about $8\%$ from $t = 20$ to $t = 50$ orbits. It increases from $t=15-20$ orbits and then fluctuates from $t = 30-50$ orbits. The intermediate and weak field cases are evolved for only $23$ orbits but show magnetic flux variation of about $5\%$ over $18$ orbits. This behavior has also been reported in \citet{ZhuStone}, where the authors found a steady or quasi-steady state balance of advection and diffusion of the magnetic field. Fig.~\ref{fig:streamline} shows that the field is advecting inward in accreting regions and being carried outward in the disk midplane.

Overall, our results indicate that the flux is neither accumulating at the center and approaching a MAD state, nor escaping from the inner disk through diffusion.  Within the limited context of our specific setup, this suggests that magnetically elevated accretion may be long-lived once it is established.
%================================================
\subsection{Comparison with Local Shearing-box Simulations}
\label{sec:localsim}
\begin{figure}
\centering
\includegraphics[width=\columnwidth]{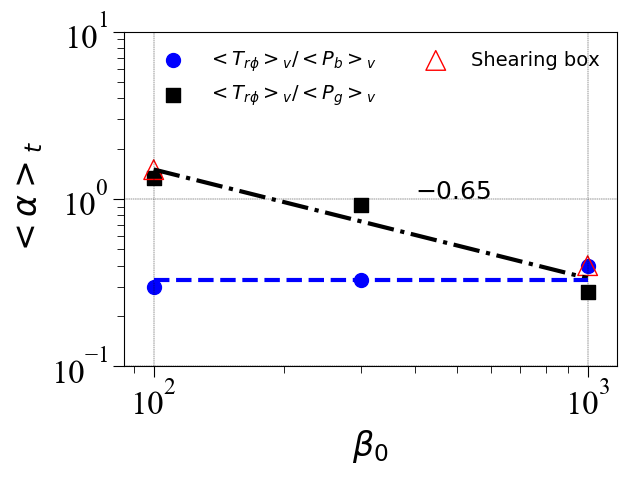}
\caption{Black points show the time averaged (19-23 orbits) viscosity parameter $\alpha=\langle T_{\mathrm{r}\phi}\rangle_v/\langle P_\mathrm{g}\rangle_v$ vs. initial $\beta = \beta_0$. The red triangles are reproduced data points from local shearing-box simulations in \citet{Greg16}. The dashed-dot fit to black data points gives a slope of $-0.65$. The blue points correspond to the analogous ratio of stress to magnetic pressure.  The dashed blue curve is almost horizontal, also consistent with the results of \citet{Greg16}.}
\label{fig:alpha_beta}
\end{figure}

As this is one of the first sets of global simulations to probe the strongly magnetized regime, it behooves us to compare our results with previous studies of strongly magnetized accretion, carried out within the local, shearing box approximation. 
In Fig.~\ref{fig:alpha_beta}, we compute the correlation of the viscosity parameter $\alpha$ with the initial $\beta_0$. If we fit a line to the three data points from our simulations in the $\log\beta_0 - \log \alpha$ plane, we find a slope of $-0.65$, which is similar to the value (-0.53) reported for local shearing box simulations \citep{Greg16}. If we compute an analogous viscosity parameter taking magnetic pressure in the denominator of Eq.~\ref{eq:alpha}, the slope in the $\log\beta_0 - \log \alpha$ plot is approximately zero, indicating that $\alpha$ in strongly magnetized disks remains unchanged if computed using magnetic pressure only.

As discussed earlier, we find that $\alpha$ values themselves (at a given $\beta_0$) are in approximate agreement between local and global simulations.  Furthermore, the power law slope of the $\alpha$--$\beta_0$ fit is in broad agreement with the results of \cite{Greg16}, though it is slightly steeper. 

Figure~\ref{fig:alpha_beta} also displays the stress normalized by the magnetic pressure, i.e., $\alpha_{\rm mag}$ \citep[see][]{Greg16}.  As with the standard $\alpha$, we find excellent agreement in the value of $\alpha_{\rm mag}$, which is $\sim 0.2$--0.4 in both \citet{Greg16} and this work.   We also find essentially no (or very weak) dependence of $\alpha_{\rm mag}$ on $\beta_0$, consistent with local studies \citep{Greg16}. 

Upon examining the vertical profiles of relevant quantities (Fig.~\ref{fig:betarhovr}), we find general qualitative agreement between local and global simulations.  In particular, gas density, pressure, and magnetic energy fall off quite rapidly away from the mid-plane for all $\beta_0$ values. As $\beta_0$ decreases, the disk column at $R = 1$ tends towards being magnetically dominated, $\langle\beta_t\rangle_{\phi} < 1$, with only a very small region of subthermal magnetic field at the midplane in our global simulations (compared with no such region in local simulations).

One exception to this agreement is in the vertical profile of the Maxwell stress, as shown in Fig.~\ref{fig:betarhovr}, which depicts both the coherent large scale stress and the turbulent stress in all of our simulations. As shown, the laminar stresses dominate nearly everywhere, apart from the midplane where the turbulent stresses dominate.  This result can be contrasted with similar profiles in \cite{Bai13} (see their Fig. 6), which shows that for $\beta_0 = 100$ and $\beta_0 = 1000$, the turbulent stress never dominates. 

Presumably, this difference in the stress profile arises from the presence of a current sheet at the midplane in our simulations.  Due to the symmetry of the shearing box, as discussed in detail in, e.g.,  \citet{Bai13}, such current sheets need not be present in local domains. However, in global setups, these current sheets act to dissipate large scale magnetic fields, which can lead to an increase in turbulent stresses at the mid-plane over laminar stresses. 

In addition to the vertical stress profile, our global simulations differ from local simulations when considering radial and azimuthal structure, particularly in the magnetic field.  While both types of simulations exhibit very strong inhomogeneities within the domain \citep{Greg16}, local simulations are by definition unable to generate the coherent large-scale spiral structures that we see here.  

In summary, there are both important similarities between local and global simulations, including the empirical relation between $\alpha$ and $\beta_0$, together with a number of results, largely related to the behavior of large-scale magnetic fields, that differ between the two approaches.  These differences are not surprising given the obvious limitations and inherent symmetries of the the shearing box. They indicate that when probing the strongly magnetized limit of accretion, local shearing boxes should be treated with caution. 
%===================================================
%=======================================
\subsection{Comparison with previous global simulations}
\label{sec:globalsim}
The closest comparison we can make to prior global simulations is with the work of \citet{ZhuStone}, who considered disks with $\beta = 10^4$ and $1000$ for two different disk scale heights of $H/r = 0.1$ and $0.05$. We also performed a $\beta=1000$ case to compare with results reported in \citet{ZhuStone}. The simulations reported here use the same code, essentially identical numerical methods, and a similar initial setup. The main distinction (apart from our consideration of lower initial $\beta$ values, leading to more strongly magnetized accretion flows) is that we adopt a disk profile with a radially constant $H/r$. We also model the full $2 \pi$ in azimuth, whereas \citet{ZhuStone} used a restricted azimuthal domain for their thin disk case. In the parameter regime where the two studies overlap, we find generally good agreement as to the predicted structure of the accretion flow. In particular, \citet{ZhuStone} and \citet{Zhu19} also found that the elevated accreting regions have stress dominated by coherent stress, whereas turbulent stress dominates near the midplane. Winds are sub-dominant as a driver of accretion in both sets of simulations.

Reducing the initial $\beta$ below the value of $10^3$ considered by \citet{ZhuStone} results in stronger disk magnetization, and some qualitatively different results. The radial structure of strongly magnetized flows shows spiral structures which were not prominent in the weakly magnetized case. The vertical structure also shows much higher density at higher altitudes and enhanced elevated accretion flow on the disk's surface layers. We also find a much higher viscosity parameter $\alpha$ due to our strong seed field. On the other hand, the result that accretion in the elevated regions is driven primarily by a coherent, rather than turbulent, Maxwell stress, persists across the weak-to-strong magnetization transition. More strongly magnetized disks show a larger mass loss rate due to winds, although we find that the mass and angular momentum loss due to the wind remain small even in the most highly magnetized case.

\citet{ZhuStone} briefly mention that a numerical simulation with $H/r = 0.1$ and $\beta_0 = 100$ was attempted but due to rapid mass loss the simulation gave very different outcomes to their weak field cases. We carried out our strong field run with $H/r = 0.05$ for $50$ orbits without substantial mass loss (only $0.3\%$ of total mass is lost during the entire run for the strong field case). It is not clear whether the important difference here is due to the different values of the aspect ratio, or to detailed differences in the numerical setup between the two sets of simulations.
%================================================
%================================================
\section{Summary and Discussion}
\label{sec:discussion}
We have presented global numerical simulations of idealized accretion disk flows threaded by weak but dynamically important net poloidal magnetic fields. Our runs span a range in the initial ratio of midplane gas to poloidal magnetic pressure between $10^3$ and $10^2$, which corresponds to the locally predicted transition between weakly and strongly magnetized disks. Using Athena++ with three levels of static mesh refinement, we were able to attain a resolution of approximately 26 cells per disk scale height in a fully global model with $H/r = 0.05$. We find that:
\begin{itemize}    
    \item The simulated disk becomes magnetic pressure supported at higher altitudes with accretion mainly occurring in the elevated regions at $z/R \approx 0.2$. Accretion in these surface layers is mainly driven by the coherent component of Maxwell stress rather than the turbulent component. The disk midplane, on the other hand, exhibits radially outward flow and is dominated by turbulent Maxwell stress. Winds are present but are not the dominant driver of accretion.
    
    \item The viscosity parameter $\alpha$ (incorporating both laminar and turbulent stress) is approximately 10 in the elevated accreting regions of the strong magnetic field case. In the disk midplane, we find an $\alpha\approx 0.5$ for the strong field case. The intermediate and weak magnetic field cases give an $\alpha$ value close to unity in the elevated regions and $\alpha < 1$ close to disk midplane. A comparison of the inferred viscosity parameter $\alpha$ from our numerical simulations and previously performed local shearing box simulations \citep{Greg16} gives similar results.
    
    \item In addition to elevated accreting regions, the disk shows spiral structures, characterized by large density inhomogeneities, that can be modeled approximately as logarithmic spirals. The spiral structures are more open as the initial magnetic field strength increases. 
    
    \item We find a quasi-steady state of the magnetic flux in all three simulations reported in this paper, suggesting that advection and diffusion of flux achieve a rough balance. 
\end{itemize}

Our results, taken together with earlier studies, suggest that the qualitative structure of accretion flows is strongly influenced by the net magnetic flux threading the disk. The net flux may be almost as important as the accretion rate, or the efficiency of cooling, in determining observable properties of the flow. All of the three possible regimes of magnetization --- effectively zero net flux, the magnetically elevated disks considered in this paper, and magnetically arrested disks --- can plausibly be realized in astrophysically relevant situations. Magnetically elevated disks may be good candidates for systems where the basic geometry of accretion appears to be that of a geometrically thin disk, but where in detail the predictions of standard thin disk models fail. We have not constructed full physics models of magnetically elevated disks (which would have to include the physics of radiation pressure and radiative transport), but it is highly plausible that such disks would be thicker, with faster inflow speeds and enhanced thermal and viscous stability, compared to standard models. They would also have decreased self-gravity and highly inhomogeneous density structures. Qualitatively, these differences go in the right direction to address longstanding problems with standard accretion disk models.

We have presented evidence that the net poloidal fields necessary for sustaining magnetically elevated disks may attain a stable equilibrium (i.e., not run away toward either a zero net flux or MAD state). Nonetheless, how the required net flux is established at different radii in the disk remains an open question. The most straightforward source for this flux would be the material feeding the flow. However, it is not clear under what conditions a thin disk starting with a weaker net flux (say, $\beta_0 > 10^4$), would be able to overcome magnetic field diffusion according to the criterion proposed in \citet{Lubow94}. Alternatively, stochastic processes could lead to the emergence of patches of net flux in the form of large fluctuations \citep{Beckwith11,Begelman14}. Longer duration numerical simulations, and simplified models for the evolution of the net flux, are needed to explore such issues.

\section*{Acknowledgements}
BM thanks to Zhaohuan Zhu for disscussions at early stages of the project. We acknowledge computational support from the PROMETHEUS supercomputer in the PL-Grid infrastructure in Poland, and acknowledge support from NASA Astrophysics Theory Program grants NNX16AI40G and NNX17AK55G. PJA acknowledges support from NASA Astrophysics Theory Program grant 80NSSC18K0640.
%===========================================================================================
%===========================================================================================
\bibliographystyle{mnras}
\bibliography{ref}

%===========================================================================================
%===========================================================================================
\end{document}